\documentclass[a4paper,USenglish]{lipics}

\usepackage{todonotes}
\usepackage{import}

\usepackage{paralist}
\usepackage{aliascnt}

\usepackage[T1]{fontenc}  

\usepackage[capitalize]{cleveref}
\usepackage[firstinits=true,maxnames=6,backend=bibtex]{biblatex}

\title{Computing downward closures for stacked counter automata}
\author{Georg Zetzsche}
\affil{AG Concurrency Theory \\Fachbereich Informatik\\TU Kaiserslautern\\\texttt{zetzsche@cs.uni-kl.de}}
\Copyright{Georg Zetzsche}

\newcommand{\rev}[1]{{#1}^R}

\newcommand{\emptyWord}{\varepsilon}
\newcommand{\Powerset}[1]{\mathcal{P}({#1})}
\newcommand{\TrivialMonoid}{\mathbf{1}}
\newcommand{\HG}{\mathsf{G}}
\newcommand{\HF}{\mathsf{F}}

\newcommand{\HomSLI}[1]{\mathsf{SLI}(#1)}
\newcommand{\Alg}[1]{\mathsf{Alg}(#1)}

\newcommand{\SententialForms}[1]{\mathsf{SF}(#1)}

\newcommand{\Dclosure}[1]{#1\mathord{\downarrow}}
\newcommand{\Uclosure}[2][YYY]{\ifthenelse{\equal{#1}{YYY}}{#2\mathord{\uparrow}}{#2\mathord{\uparrow}_{#1}}}
\newcommand{\Parikh}[1]{\Psi(#1)}
\newcommand{\ParikhInv}[1]{\Psi^{-1}(#1)}
\newcommand{\ParikhMap}{\Psi}
\newcommand{\Lang}[1]{\mathsf{L}(#1)}

\newcommand{\yield}[1]{\mathsf{yield}(#1)}

\newcommand{\StCtr}{\mathsf{SC}}

\newcommand{\M}[1][]{\mathbb{M}\ifthenelse{\equal{#1}{}}{}{(#1)}}

\newcommand{\ratunion}{\cup}

\newcommand{\congruence}{\equiv}

\newcommand{\C}{\mathcal{C}}

\newcommand{\B}{\mathbb{B}}

\newcommand{\RInv}[2][YYY]{\ifthenelse{\equal{#1}{YYY}}{\overrightarrow{\mathsf{I}}(#2)}{\overrightarrow{\mathsf{I}}_{#1}(#2)}}
\newcommand{\LInv}[2][YYY]{\ifthenelse{\equal{#1}{YYY}}{\overleftarrow{\mathsf{I}}(#2)}{\overleftarrow{\mathsf{I}}_{#1}(#2)}}

\newcommand{\Rclass}{R}

\newcommand{\VA}[1]{\mathsf{VA}(#1)}

\newcommand{\Reg}{\mathsf{REG}}
\newcommand{\CF}{\mathsf{CF}}

\newcommand{\AH}[1][XX]{\mathsf{AH}\ifthenelse{\equal{#1}{XX}}{}{(#1)}}
\newcommand{\RE}[1][XX]{\mathsf{RE}\ifthenelse{\equal{#1}{XX}}{}{(#1)}}
\newcommand{\REC}[1][XX]{\mathsf{REC}\ifthenelse{\equal{#1}{XX}}{}{(#1)}}

\newcommand{\defeq}{=}

\newcommand{\autstep}[1][YYY]{\ifthenelse{\equal{#1}{YYY}}{\rightarrow}{\rightarrow_{#1}}}
\newcommand{\autsteps}[1][YYY]{\ifthenelse{\equal{#1}{YYY}}{\rightarrow^*}{\rightarrow^*_{#1}}}
\newcommand{\autstepsn}[2][YYY]{\ifthenelse{\equal{#1}{YYY}}{\rightarrow^{#2}}{\rightarrow^{#2}_{#1}}}
\newcommand{\grammarstep}[1][YYY]{\ifthenelse{\equal{#1}{YYY}}{\Rightarrow}{\Rightarrow_{#1}}}
\newcommand{\grammarsteps}[1][YYY]{\ifthenelse{\equal{#1}{YYY}}{\Rightarrow^*}{\Rightarrow^*_{#1}}}
\newcommand{\grammarstepsn}[2][YYY]{\ifthenelse{\equal{#1}{YYY}}{\Rightarrow^{#2}}{\Rightarrow^{#2}_{#1}}}

\newcommand{\N}{\mathbb{N}}

\newcommand{\Z}{\mathbb{Z}}

\theoremstyle{plain}
\newtheorem{ctheorem}{Theorem}
\crefname{ctheorem}{Theorem}{Theorems}
\Crefname{ctheorem}{Theorem}{Theorems}

\newaliascnt{clemma}{ctheorem}		
\newtheorem{clemma}[clemma]{Lemma}	
\aliascntresetthe{clemma}		
\crefname{clemma}{Lemma}{Lemmas}
\Crefname{clemma}{Lemma}{Lemmas}

\newaliascnt{cdefinition}{ctheorem}
\newtheorem{cdefinition}[cdefinition]{Definition}
\aliascntresetthe{cdefinition}
\crefname{cdefinition}{Definition}{Definitions}
\Crefname{cdefinition}{Definition}{Definitions}

\newaliascnt{cproposition}{ctheorem}
\newtheorem{cproposition}[cproposition]{Proposition}
\aliascntresetthe{cproposition}
\crefname{cproposition}{Proposition}{Propositions}
\Crefname{cproposition}{Proposition}{Propositions}

\newaliascnt{ccorollary}{ctheorem}

\aliascntresetthe{ccorollary}
\crefname{ccorollary}{Corollary}{Corollarys}
\Crefname{ccorollary}{Corollary}{Corollaries}

\newaliascnt{cexample}{ctheorem}
\newtheorem{cexample}[cexample]{Example}
\aliascntresetthe{cexample}
\crefname{cexample}{Example}{Examples}
\Crefname{cexample}{Example}{Examples}

\def\shuf{\mathbin{\mathchoice
{\rule{.03em}{1ex}\rule{.3em}{.03em}\rule{.03em}{1ex}
 \rule{.3em}{.03em}\rule{.03em}{1ex}}
{\rule{.03em}{1ex}\rule{.3em}{.03em}\rule{.03em}{1ex}
 \rule{.3em}{.03em}\rule{.03em}{1ex}}
{\rule{.02em}{.7ex}\rule{.2em}{.02em}\rule{.2pt}{.7ex}
 \rule{.2em}{.02em}\rule{.02em}{.7ex}}
{\rule{.03em}{1ex}\rule{.3em}{.03em}\rule{.03em}{1ex}
 \rule{.3em}{.03em}\rule{.03em}{1ex}}
}}

\newcommand{\shuffle}{\shuf}

\begin{document}
\maketitle

\begin{abstract}
The downward closure of a language $L$ of words is the set of all (not
necessarily contiguous) subwords of members of $L$. It is well known that the
downward closure of any language is regular. Although the downward
closure seems to be a promising abstraction, there are only few language
classes for which an automaton for the downward closure is known to be
computable.

It is shown here that for stacked counter automata, the downward closure is
computable.  Stacked counter automata are finite automata with a storage
mechanism obtained by \emph{adding blind counters} and \emph{building stacks}.
Hence, they generalize pushdown and blind counter automata.

The class of languages accepted by these automata are precisely those in the
hierarchy obtained from the context-free languages by alternating two closure
operators: imposing semilinear constraints and taking the algebraic extension.
The main tool for computing downward closures is the new concept of Parikh
annotations.  As a second application of Parikh annotations, it is shown that
the hierarchy above is strict at every level.
\end{abstract}

\section{Introduction}

In the analysis of systems whose behavior is given by formal languages, it is a
fruitful idea to consider abstractions: simpler objects that preserve relevant
properties of the language and are amenable to algorithmic examination.  A very
well-known such type of abstraction is the \emph{Parikh image}, which counts
the number of occurrences of each letter. For a variety of language classes,
the Parikh image of every language is known to be effectively semilinear, which
facilitates a range of analysis techniques for formal
languages (see \cite{KopczynskiTo2010} for applications).

A promising alternative to Parikh images is the \emph{downward closure}
$\Dclosure{L}$, which consists of all (not necessarily contiguous) subwords of
members of $L$. Whereas for many interesting classes of languages the Parikh
image is not semilinear in general, the downward closure is regular \emph{for
any language}, suggesting wide applicability. Moreover, the downward closure
encodes properties not visible in the Parikh image: Suppose $L$ describes the
behavior of a system that is observed through a lossy channel, meaning that on
the way to the observer, arbitrary actions can get lost. Then, $\Dclosure{L}$
is the set of words received by the observer~\cite{HabermehlMeyerWimmel2010}.
Hence, given the downward closure as a finite automaton, we can decide whether
two systems are equivalent under such observations, and even whether the
behavior of one system includes the other.  Hence, even if Parikh images
are effectively semilinear for a class of languages, computing the downward
closure is still an important task.
See~\cite{AtigBouajjaniQadeer2009,LongCalinMajumdarMeyer2012} for further
applications.

However, while there always \emph{exists} a finite automaton for the downward
closure, it seems difficult to \emph{compute} them and there are few language
classes for which computability has been established. The downward closure is
computable for context-free languages and algebraic
extensions~\cite{vanLeeuwen1974,Courcelle1991}, backward reachability sets of
lossy channel systems~\cite{AbdullaJonsson1996}, 0L-systems and context-free
FIFO rewriting systems~\cite{AbdullaBoassonBouajjani2001}, and Petri net
languages~\cite{HabermehlMeyerWimmel2010}.  It is not computable
for reachability sets of lossy channel systems~\cite{Mayr2003} and for
Church-Rosser languages~\cite{GruberHolzerKutrib2007}.

It is shown here that downward closures are computable for \emph{stacked
counter automata}.  These are automata with a finite state control and a
storage mechanism obtained by two constructions (of storage mechanisms): One
can \emph{build stacks} and \emph{add blind counters}. The former is to
construct a new mechanism that stores a stack whose entries are configurations
of an old mechanism. One can then manipulate the topmost entry, pop it if
empty, or start a new one on top.  Adding a blind counter to an old mechanism
yields a new mechanism in which the old one and a blind counter (i.e., a
counter that can attain negative values and has to be zero in the end of a run)
can be used simultaneously.  

Stacked counter automata are interesting because among a large class of
automata with storage, they are \emph{expressively complete} for those storage
mechanisms that guarantee semilinear Parikh images.  This is due to the fact
that they accept precisely those languages in the hierarchy obtained from the
context-free languages by alternating two closure operators: imposing
semilinear constraints (with respect to the Parikh image) and taking the
algebraic extension. These two closure operators correspond to the
constructions of storage mechanisms in stacked counter automata
(see~\cref{sec:hierarchy}).

The main tool to show the computability of downward closures is the concept of
\emph{Parikh annotations}.  As another application of this concept, it is shown
that the aforementioned hierarchy is strict at every level. 

The paper is structured as follows. After \cref{sec:preliminaries} defines
basic concepts and notation, \cref{sec:hierarchy} introduces the hierarchy of
language classes.  \Cref{sec:pa} presents Parikh annotations, the main
ingredient for the computation of downward closures. The main result is then
presented in \cref{sec:dclosure}, where it is shown that downward closures are
computable for stacked counter automata. As a second application of Parikh
annotations, it is then shown in \cref{sec:strictness} that the hierarchy
defined in \cref{sec:hierarchy} is strict at every level.  Unfortunately, due
to space restrictions, most proofs had to be moved to the appendix.

\section{Preliminaries}
\label{sec:preliminaries}
A \emph{monoid} is a set $M$ together with a binary associative operation such
that $M$ contains a neutral element. Unless the monoid at hand warrants a
different notation, we will denote the neutral element by $1$ and the product
of $x,y\in M$ by $xy$. The trivial monoid that contains only the neutral element is
denoted by $\TrivialMonoid$. 

If $X$ is an alphabet, $X^*$ denoted the set of words over $X$.  The
empty word is denoted by $\emptyWord\in X^*$.  For a symbol $x\in X$ and a word
$w\in X^*$, let $|w|_x$ be the number of occurrences of $x$ in $w$ and
$|w|=\sum_{x\in X}|w|_x$.  For an alphabet $X$ and languages $L,K\subseteq
X^*$, the \emph{shuffle product} $L\shuffle K$ is the set of all words $u_0 v_1
u_1 \cdots v_n u_n$ where $u_0,\ldots,u_n,v_1,\ldots,v_n\in X^*$, $u_0\cdots
u_n\in L$, and $v_1\cdots v_n\in K$.  For a subset $Y\subseteq X$, we define
the \emph{projection morphism} $\pi_Y\colon X^*\to Y^*$ by $\pi_Y(y)=y$ for
$y\in Y$ and $\pi_Y(x)=\emptyWord$ for $x\in X\setminus Y$.  By $\Powerset{S}$,
we denote the power set of the set $S$. A \emph{substitution} is a map
$\sigma\colon X\to \Powerset{Y^*}$ and given $L\subseteq X^*$, we write
$\sigma(L)$ for the set of all words $v_1\cdots v_n$, where $v_i\in
\sigma(x_i)$, $1\le i\le n$, for $x_1\cdots x_n\in L$ and $x_1,\ldots,x_n\in
X$. If $\sigma(x)\subseteq Y$ for each $x\in X$, we call $\sigma$ a
\emph{letter substitution}.

For words $u,v\in X^*$, we write $u\preceq v$ if $u=u_1\cdots u_n$ and
$v=v_0u_1v_1\cdots u_nv_n$ for some $u_1,\ldots,u_n,v_0,\ldots,v_n\in X^*$.  It
is well-known that $\preceq$ is a well-quasi-order on $X^*$ and that therefore
the \emph{downward closure} $\Dclosure{L}=\{u\in X^* \mid \exists v\in L\colon
u\preceq v\}$ is regular for any $L\subseteq X^*$~\cite{Higman1952}.

If $X$ is an alphabet, $X^\oplus$ denotes the set of maps $\alpha\colon
X\to\N$.  The elements of $X^\oplus$ are called \emph{multisets}.  Let
$\alpha+\beta\in X^\oplus$ be defined by
$(\alpha+\beta)(x)=\alpha(x)+\beta(x)$.  With this operation, $X^\oplus$ is a
monoid.  We consider each $x\in X$ to be an element of $X^\oplus$.  For a
subset $S\subseteq X^\oplus$, we write $S^\oplus$ for the smallest submonoid of
$X^\oplus$ containting $S$. For $\alpha\in X^\oplus$ and $k\in\N$, we define
$(k\cdot \alpha)(x)=k\cdot\alpha(x)$, meaning $k\cdot\alpha\in X^\oplus$. A subset of the form 
$\mu+F^\oplus$
for $\mu\in X^\oplus$ and a finite $F\subseteq X^\oplus$ is called
\emph{linear}. A finite union of linear sets is called \emph{semilinear}.  The
\emph{Parikh map} is the map $\ParikhMap\colon X^*\to X^\oplus$ defined by
$\Parikh{w}(x)\defeq |w|_x$ for all $w\in X^*$ and $x\in X$.  Given a morphism
$\varphi\colon X^\oplus\to Y^\oplus$ and a word $w\in X^*$, we use $\varphi(w)$
as a shorthand for $\varphi(\Parikh{w})$.  We lift $\ParikhMap$ to
sets in the usual way: $\Parikh{L}=\{\Parikh{w} \mid w\in L\}$. If
$\Parikh{L}$ is semilinear, we will also call $L$ itself semilinear.

Let $M$ be a monoid. An \emph{automaton over $M$} is a tuple $A=(Q,M,E,q_0,F)$,
in which
\begin{inparaenum}[(i)]
\item $Q$ is a finite set of \emph{states},
\item $E$ is a finite subset of $Q\times M\times Q$ called the set of \emph{edges},
\item $q_0\in Q$ is the \emph{initial state}, and
\item $F\subseteq Q$ is the set of \emph{final states}.
\end{inparaenum}
We write $(q,m) \autstep[A] (q',m')$ if there is an edge
$(q,r,q')\in E$ such that $m'=mr$. The set \emph{generated} by $A$ is then
$S(A)\defeq\{m\in M \mid (q_0,1)\autsteps[A] (f,m)~\text{for some $f\in F$} \}$.

A \emph{finite state transducer} is an automaton over $Y^*\times X^*$ for
alphabets $X,Y$.  Relations of the form $S(A)$ for finite state
transducers $A$ are called \emph{rational transductions}.  For a language
$L\subseteq X^*$ and a rational transduction $T\subseteq Y^*\times X^*$, we
write $TL=\{u\in Y^* \mid \exists v\in L\colon  (u,v)\in T\}$. If $TF$ is
finite for every finite language $F$, $T$ is said to be \emph{locally finite}.
A class $\C$ of languages is called a \emph{full trio} if it is closed under
rational transductions, i.e. if $TL\in\C$ for every $L\in\C$ and every rational
transduction $T$. It is called a \emph{full semi-trio} if it is closed under
locally finite rational transductions. A full \emph{semi-AFL} is a union closed
full trio. 

\subparagraph{Stacked counter automata} In order to define stacked counter
automata, we use the concept of valence automata, which combine a finite state
control with a storage mechanism defined by a monoid $M$.  A \emph{valence
automaton over $M$} is an automaton $A$ over $X^*\times M$ for an alphabet $X$.
The \emph{language accepted by $A$} is then $\Lang{A}=\{w\in X^* \mid (w,1)\in
S(A)\}$. The class of languages accepted by valence automata over $M$ is
denoted $\VA{M}$. By choosing suitable monoids $M$, one can obtain various
kinds of automata with storage as valence automata.  For example, blind
counters, partially blind counters, pushdown storages, and combinations thereof
can all be realized by appropriate monoids~\cite{Zetzsche2013a}.

If one storage mechanism is realized by a monoid $M$, then the mechanism that
\emph{builds stacks} is realized by the monoid $\B*M$. Here, $\B$ denotes the
bicyclic monoid, presented by $\langle a,\bar{a} \mid a\bar{a}=1\rangle$, and
$*$ denotes the free product of monoids. For readers not familiar with these
concepts, it will suffice to know that a configuration of the storage mechanism
described by $\B*M$ consists of a sequence $c_0ac_1\cdots ac_n$, where
$c_0,\ldots,c_n$ are configurations of the mechanism realized by $M$.  We
interpret this as a stack with the entries $c_0,\ldots,c_n$.  One can open a
new stack entry on top (by multiplying $a\in\B$), remove the topmost entry if
empty (by multiplying $\bar{a}\in\B$) and operate on the topmost entry  using
the old mechanism (by multiplying elements from $M$).  For example, the monoid
$\B$ describes a partially blind counter (i.e. a counter that cannot go below
zero and is only tested for zero in the end) and $\B*\B$ describes a pushdown with two stack
symbols. Given a storage mechanism realized by a monoid $M$, we can \emph{add a
blind counter} by using the monoid $M\times\Z$, where $\Z$ denotes the group of
integers.  We define $\StCtr$ to be the smallest class of monoids 
with $\TrivialMonoid\in\StCtr$ such that whenever $M\in\StCtr$, we also have
$M\times\Z\in\StCtr$ and $\B*M\in\StCtr$.  A \emph{stacked counter automaton} is a
valence automaton over $M$ for some $M\in\StCtr$. For more details,
see~\cite{Zetzsche2013a}.  In \cref{sec:hierarchy}, we will turn to a different
description of the languages accepted by stacked counter automata.

\section{A hierarchy of language classes}
\label{sec:hierarchy}
This section introduces a hierarchy of language classes that divides the class
of languages accepted by stacked counter automata into levels. This will allow
us to apply recursion with respect to these levels.  
The hierarchy is defined by alternating two operators on language classes,
algebraic extensions and semilinear intersections.

\subparagraph*{Algebraic extensions}
Let $\C$ be a class of languages. A \emph{$\C$-grammar} is a quadruple
$G=(N,T,P,S)$ where $N$ and $T$ are disjoint alphabets and $S\in N$.  The
symbols in $N$ and $T$ are called the \emph{nonterminals} and the
\emph{terminals}, respectively.  $P$ is a finite set of pairs $(A,M)$ with
$A\in N$ and $M\subseteq (N\cup T)^*$, $M\in\C$.  A pair $(A,M)\in P$ is called
a \emph{production of $G$} and also denoted by $A\to M$.  The set $M$ is the
\emph{right-hand side} of the production $A\to M$. 

We write $x\grammarstep[G] y$ if $x=uAv$ and $y=uwv$ for some $u,v,w\in (N\cup
T)^*$ and $(A,M)\in P$ with $w\in M$. A word $w$ with $S\grammarstep[G]^* w$ is
called a \emph{sentential form} of $G$ and we write $\SententialForms{G}$ for
the set of sentential forms of $G$.  The \emph{language generated by $G$} is
$\Lang{G}=\SententialForms{G}\cap T^*$.  Languages generated by $\C$-grammars
are called \emph{algebraic over $\C$}.  The class of all languages that are
algebraic over $\C$ is called the \emph{algebraic extension} of $\C$ and
denoted $\Alg{\C}$. We say a language class $\C$ is \emph{algebraically closed}
if $\Alg{\C}=\C$.  If $\C$ is the class of finite languages, $\C$-grammars
are also called \emph{context-free grammars}.

We will use the operator $\Alg{\cdot}$ to describe the effect of \emph{building
stacks} on the accepted languages of valence automata.
In~\cite{Zetzsche2013a}, it was shown that $\VA{M_0 * M_1}\subseteq\Alg{\VA{M_0}\cup\VA{M_1}}$.  Here, we complement this by showing that if one
of the factors is $\B*\B$, the inclusion becomes an equality.
Observe that since $\VA{\B*\B}$ is the class of languages accepted by pushdown
automata and $\Alg{\Reg}=\Alg{\VA{\TrivialMonoid}}$ is clearly the class of
languages generated by context-free grammars, the first statement of the
following \lcnamecref{algbb} generalizes the equivalence between pushdown
automata and context-free grammars.
\begin{ctheorem}\label{algbb}
For every monoid $M$, $\Alg{\VA{M}}=\VA{\B*\B*M}$. 
\end{ctheorem}

\subparagraph*{Semilinear intersections}
The second operator on language classes lets us describe the languages in
$\VA{M\times\Z^n}$ in terms of those in $\VA{M}$. Consider a
language class $\C$. By $\HomSLI{\C}$, we denote the class of languages of the
form $h(L\cap\ParikhInv{S})$, where $L\subseteq X^*$ is in $\C$, the set
$S\subseteq X^\oplus$ is semilinear, and $h\colon X^*\to Y^*$ is a morphism.
We call a language class $\C$ \emph{Presburger closed} if $\HomSLI{\C}=\C$.
The following proof requires only standard techniques.
\begin{cproposition}\label{basics:sli:zpowers}
Let $M$ be a monoid. Then $\HomSLI{\VA{M}}=\bigcup_{n\ge 0} \VA{M\times\Z^n}$.
\end{cproposition}

The hierarchy is now obtained by alternating the operators $\Alg{\cdot}$ and $\HomSLI{\cdot}$.
Let $\HF_0$ be the class of finite languages and let
\begin{align*}
\HG_i=\Alg{\HF_i},~~~~\HF_{i+1}=\HomSLI{\HG_i}~~~\text{for each $i\ge 1$}, && \HF=\bigcup_{i\ge 0} \HF_i.
\end{align*}
Then we clearly have the inclusions
$\HF_0\subseteq\HG_0\subseteq\HF_1\subseteq\HG_1\subseteq\cdots$.  Furthermore,
$\HG_0$ is the class of context-free languages, $\HF_1$ is the smallest
Presburger closed class containing $\CF$, $\HG_1$ the algebraic extension of
$\HF_1$, etc. In particular, $\HF$ is the smallest Presburger closed and
algebraically closed language class containing the context-free languages.

The following \lcnamecref{basics:fsemilinear} is due to the fact that both
$\Alg{\cdot}$ and $\HomSLI{\cdot}$ preserve (effective) semilinearity. The
former has been shown by van~Leeuwen~\cite{vanLeeuwen1974}.
\begin{cproposition}\label{basics:fsemilinear}
The class $\HF$ is effectively semilinear.
\end{cproposition}

The work~\cite{BuckheisterZetzsche2013a} characterized all those storage
mechanisms among a large class (namely among those defined by graph products of
the bicyclic monoid and the integers) that guarantee semilinear Parikh images.
Each of the corresponding language classes was obtained by alternating the
operators $\Alg{\cdot}$ and $\HomSLI{\cdot}$, meaning that all these classes
are contained in $\HF$. Hence, the following means that stacked counter
automata are expressively complete for these storage mechanisms. It follows
directly from \cref{algbb,basics:sli:zpowers}.
\begin{ctheorem}\label{basics:stctr:f}
Stacked counter automata accept precisely the languages in $\HF$.
\end{ctheorem}

One might wonder why $\HF_0$ is not chosen to be the regular languages. While
this would be a natural choice, our recursive algorithm for computing downward closures
relies on the following fact
.
Note that the regular languages are not Presburger closed.
\begin{cproposition}\label{hierarchy:closure}
For each $i\ge 0$, the class $\HF_i$ is an effective Presburger closed full
semi-trio.  Moreover, for each $i\ge 0$, $\HG_i$ is an effective full semi-AFL.
\end{cproposition}

\section{Parikh annotations}
\label{sec:pa}

\newcommand{\marker}{\diamond}

This section introduces Parikh annotations, the key tool in our procedure for
computing downward closures. Suppose $L$ is a semilinear language.  Then for
each $w\in L$, $\Parikh{w}$ can be decomposed into a constant vector and a
linear combination of period vectors from the semilinear representation of
$\Parikh{L}$. We call such a decomposition a \emph{Parikh decomposition}.  The
main purpose of Parikh annotations is to provide transformations of languages
that \emph{make reference to Parikh decompositions} without leaving the
respective language class. For example, suppose we want to transform a
context-free language $L$ into the language $L'$ of all those words $w\in L$
whose Parikh decomposition does not contain a specified period vector.  
This may not be possible with rational transductions: If $L_\vee=\{a^nb^m \mid
\text{$m=n$ or $m=2n$}\}$, then the Parikh image is $(a+b)^\oplus \cup
(a+2b)^\oplus$, but a finite state transducer cannot determine whether the
input word has a Parikh image in $(a+b)^\oplus$ or in $(a+2b)^\oplus$.
Therefore, a Parikh annotation for $L$ is a language $K$ in the same class with
additional symbols that allow a finite state transducer (that is applied to
$K$) to access the Parikh decomposition.

\begin{cdefinition}\label{def:parikh:annotation}
Let $L\subseteq X^*$ be a language and $\C$ be a language class. A \emph{Parikh
annotation (PA) for $L$ in $\C$} is a tuple $(K,C,P,(P_c)_{c\in C},\varphi)$, where
\begin{inparaenum}[(1)]
\item $C,P$ are alphabets such that $X,C,P$ are pairwise disjoint,
\item $K\subseteq C(X\cup P)^*$ is in $\C$,
\item $\varphi$ is a morphism $\varphi\colon (C\cup P)^\oplus\to X^\oplus$,
\item $P_c$ is a subset $P_c\subseteq P$ for each $c\in C$,
\end{inparaenum}
such that
\begin{enumerate}[(i)]
\item\label{pa:cond:proj} $\pi_X(K)=L$ (the \emph{projection property}),
\item\label{pa:cond:equality} $\varphi(\pi_{C\cup P}(w))=\Parikh{\pi_X(w)}$ for each $w\in K$ (the \emph{counting property}), and
\item\label{pa:cond:soundcomplete} $\Parikh{\pi_{C\cup P}(K)}=\bigcup_{c\in C} c+P_c^\oplus$ (the \emph{commutative projection property}).
\end{enumerate}
\end{cdefinition}
Intuitively, a Parikh annotation describes for each $w$ in $L$ one or more
Parikh decompositions of $\Parikh{w}$. The symbols in $C$ represent constant
vectors and symbols in $P$ represent period vectors. Here, the symbols in
$P_c\subseteq P$ correspond to those that can be added to the constant vector
corresponding to $c\in C$.  Furthermore, for each $x\in C\cup P$, $\varphi(x)$
is the vector represented by $x$.  The projection property states that removing
the symbols in $C\cup P$ from words in $K$ yields $L$.  The commutative
projection property requires that after $c\in C$ only symbols representing
periods in $P_c$ are allowed and that all their combinations occur.  Finally,
the counting property says that the additional symbols in $C\cup P$ indeed
describe a Parikh decomposition of $\Parikh{\pi_X(w)}$. Clearly, the conditions of a
Parikh annotation imply that $L$ is semilinear.

\begin{cexample}\label{example:pa}
Let $X=\{a,b,c,d\}$ and consider the regular set $L=(ab)^*(ca^*\ratunion db^*)$. For
$K=e(pab)^*c(qa)^* \ratunion f(rab)^*d(sb)^*$,
$P=\{p,q,r,s\}$, and $\varphi\colon (C\cup P)^\oplus\to X^\oplus$ with
$C   =\{e,f\}$,  $P_e =\{p,q\}$,  $P_f =\{r,s\}$,            $\varphi(e)=c$,    $\varphi(f)=d$, 
              $\varphi(p)=a+b$,  $\varphi(q)=a$,
              $\varphi(r)=a+b$,  and $\varphi(s)=b$,
the tuple $(K,C,P,(P_g)_{g\in C},\varphi)$ is a Parikh annotation for $L$ in $\Reg$.
\end{cexample}

In a Parikh annotation, for each $cw\in K$ and
$\mu\in P_c^\oplus$, we can find a word $cw'\in K$ such that
$\Parikh{\pi_{C\cup P}(cw')}=\Parikh{\pi_{C\cup P}(cw)}+\mu$.  In particular,
this means $\Parikh{\pi_X(cw')}=\Parikh{\pi_X(cw)}+\varphi(\mu)$.  In our
applications, we will need a further guarantee that provides such words, but
with additional information on their structure. Such a guarantee is granted by
Parikh annotations with insertion marker. Suppose $\marker\notin X$ and $u\in (X\cup\{\marker\})^*$
with $u=u_0\marker u_1\cdots\marker u_n$ for $u_0,\ldots,u_n\in
X^*$.  Then we write $u\preceq_\marker v$ if
$v=u_0v_1u_1\cdots v_nu_n$ for some $v_1,\ldots,v_n\in X^*$.
\begin{cdefinition}
Let $L\subseteq X^*$ be a language and $\C$ be a language class. A \emph{Parikh
annotation with insertion marker (PAIM) for $L$ in $\C$} is a tuple
$(K,C,P,(P_c)_{c\in C},\varphi,\marker)$ such that:
\begin{enumerate}[(i)]
\item $\marker\notin X$ and $K\subseteq C(X\cup P\cup \{\marker\})^*$ is in $\C$,
\item $(\pi_{C\cup X\cup P}(K),C,P,(P_c)_{c\in C},\varphi)$ is a Parikh annotation for $L$ in $\C$,
\item there is a $k\in\N$ such that every $w\in K$ satisfies $|w|_\marker\le k$ (\emph{boundedness}), and
\item for each $cw\in K$ and $\mu\in P_c^\oplus$, there is a $w'\in L$ with
$\pi_{X\cup\marker}(cw)\preceq_{\marker} w'$ and
$\Parikh{w'}=\Parikh{\pi_X(cw)}+\varphi(\mu)$. This property is called the
\emph{insertion property}.
\end{enumerate}
If $|C|=1$, then the PAIM is called \emph{linear} and we also write
$(K,c,P_c,\varphi,\marker)$ for the PAIM, where $C=\{c\}$. 
\end{cdefinition}
In other words, in a PAIM, each $v\in L$ has an annotation $cw\in K$ in which a
bounded number of positions is marked such that for each $\mu\in P_c^\oplus$,
we can find a $v'\in L$ with $\Parikh{v'}=\Parikh{v}+\varphi(\mu)$
such that $v'$ is obtained from $v$ by inserting words in corresponding
positions in $v$. In particular, this guarantees $v\preceq v'$.

\begin{cexample}
Let $L$ and $(K,C,P,(P_c)_{c\in C},\varphi)$ be as in \cref{example:pa}.
Furthermore, let $K' = e\marker(pab)^*c\marker(qa)^* \cup
f\marker(rab)^*d\marker(sb)^*$.  Then $(K',C,P,(P_c)_{c\in C},\varphi,\marker)$
is a PAIM for $L$ in $\Reg$.  Indeed, every word in $K'$ has at most two
occurrences of $\marker$.  Moreover, if $ew=e\marker(pab)^mc\marker (qa)^n\in
K'$ and $\mu\in P_e^\oplus$, $\mu=k\cdot p+\ell\cdot q$, then
$w'=(ab)^{k+m}ca^{\ell+n}\in L$ satisfies $\pi_{X\cup \marker}(ew) = \marker
(ab)^m c \marker a^n \preceq_\marker (ab)^k (ab)^m c a^\ell a^n=w'$ and clearly
$\Parikh{\pi_X(w')}=\Parikh{\pi_X(ew)}+\varphi(\mu)$ (and similarly for words
$fw\in K'$).
\end{cexample}

The main result of this section is that there is an algorithm that,
given a language $L\in\HF_i$ or $L\in\HG_i$, constructs a PAIM for $L$ in
$\HF_i$ or $\HG_i$, respectively.
\begin{ctheorem}\label{pa:parikhannotations}
Given $i\in\N$ and $L$ in $\HF_i$ ($\HG_i$), one can construct a
PAIM for $L$ in $\HF_i$ ($\HG_i$).
\end{ctheorem}

\subparagraph*{Outline of the proof} The rest of this section is devoted to the
proof of \cref{pa:parikhannotations}.  The construction of PAIM proceeds
recursively with respect to the level of our hierarchy.  This means, we show
that if PAIM can be constructed for $\HF_i$, then we can compute them for
$\HG_i$ (\cref{pa:alg}) and if they can be constructed for $\HG_i$, then they
can be computed for $\HF_{i+1}$ (\cref{pa:sli}). While the latter can be done
with a direct construction, the former requires a series of involved steps:
\begin{itemize}
\item The general idea is to use recursion with respect to the number of
nonterminals: Given a $\HF_i$-grammar for $L\in\HG_i$, we present $L$ in terms
of languages whose grammars use fewer nonterminals. This presentation is done
via substitutions and by using grammars with one nonterminal. 
The idea of presenting a language in $\Alg{\C}$
using one-nonterminal grammars and substitutions follows van~Leeuwen's proof of
Parikh's theorem~\cite{vanLeeuwen1974}. 
\item We construct PAIM for languages generated by one-nonterminal grammars
where we are given PAIM for the right-hand-sides (\cref{pa:onenonterminal}).
\item We construct PAIM for languages $\sigma(L)$, where $\sigma$ is a
substitution, a PAIM is given for $L$ and for each $\sigma(x)$
(\cref{pa:substitution}). This construction is again divided into the case where
$\sigma$ is a letter substitution
(i.e., one in which each symbol is mapped to a set of letters)
and the
general case. Since the case of letter substitutions constitutes the
conceptually most involved step, part of its proof is contained in this
extended abstract (\cref{consistent:substitution}).
\end{itemize}

Maybe surprisingly, the most conceptually involved step in the construction of
PAIM lies within obtaining a Parikh annotation for $\sigma(L)$ in $\Alg{\C}$,
where $\sigma$ is a letter substitution and a PAIM for $L\subseteq X^*$ in
$\Alg{\C}$ is given. This is due to the fact that one has to substitute the
symbols in $X$ consistently with the symbols in $C\cup P$; more precisely, one
has to maintain the agreement between $\varphi(\pi_{C\cup P}(\cdot))$ and
$\Parikh{\pi_X(\cdot)}$.

In order to exploit the fact that this agreement exists in the first place, we
use the following simple yet very useful \lcnamecref{nonterminal:extension}. It
states that for a morphism $\psi$ into a group, the only way a grammar $G$ can
guarantee $\Lang{G}\subseteq\psi^{-1}(h)$ is by encoding into each nonterminal
$A$ the value $\psi(u)$ for the words $u$ that $A$ derives.  The $G$-compatible
extension of $\psi$ reconstructs this value for each nonterminal.  Let
$G=(N,T,P,S)$ be a $\C$-grammar and $M$ be a monoid. A morphism $\psi\colon
(N\cup T)^*\to M$ is called \emph{$G$-compatible} if $u\grammarsteps[G] v$
implies $\psi(u)=\psi(v)$ for $u,v\in (N\cup T)^*$.  Moreover, we call $G$
\emph{reduced} if for each $A\in N$, we have $A\grammarsteps[G] w$ for some
$w\in T^*$ and $S\grammarsteps[G] uAv$ for some $u,v\in (N\cup T)^*$. 
\begin{clemma}\label{nonterminal:extension}
Let $H$ be a group and $\psi\colon T^*\to H$ be a morphism.  Furthermore, let
$G=(N,T,P,S)$ be a reduced $\C$-grammar with $\Lang{G}\subseteq\psi^{-1}(h)$
for some $h\in H$.  Then $\psi$ has a unique $G$-compatible extension
$\hat{\psi}\colon(N\cup T)^*\to H$. If $H=\Z$ and $\C=\HF_i$, $\hat{\psi}$ can
be computed.
\end{clemma}

We continue with the problem of replacing $C\cup P$ and $X$ consistently. In
order to simplify the setting and utilize the symmetry of the roles played by
$C\cup P$ and $X$, we consider a slightly more general situation. There is an
alphabet $X=X_0\uplus X_1$, morphisms $\gamma_i\colon X_i^*\to\N$, $i=0,1$, and
some $L\subseteq X^*$, $L\in\Alg{\HF_i}$ with
$\gamma_0(\pi_{X_0}(w))=\gamma_1(\pi_{X_1}(w))$ for every $w\in L$. We wish to
construct a language $L'$ in $\Alg{\HF_i}$ where each word in $L'$ is obtained
from a word in $L$ as follows. We substitute each occurrence of $x\in X_i$ by
one of $\gamma_i(x)$ many symbols $y$ in an alphabet $Y_i$, each of which will
be assigned a value $0\le \eta_i(y)\le\gamma_i(x)$. Here, we want to guarantee
that in every resulting word $w\in (Y_0\cup Y_1)^*$, we have
$\eta_0(\pi_{Y_0}(w))=\eta_1(\pi_{Y_1}(w))$, meaning that the symbols in $X_0$ and
in $X_1$ are replaced \emph{consistently}. Formally, we have 
\begin{equation} Y_i=\{(x,j) \mid x\in X_i, ~0\le j\le\gamma_i(x)\}, ~i=0,1, ~~Y=Y_0\cup Y_1\label{consistent:substitution:alphabet}\end{equation}
and the morphisms 
\begin{align}
h_i\colon Y_i^*&\longrightarrow X_i^*,   & h\colon Y^*&\longrightarrow X^*, & \eta_i\colon Y_i^* &\longrightarrow\N,    \label{consistent:substitution:morphisms} \\
(x,j)&\longmapsto x,                     & (x,j) &\longmapsto x,            & (x,j)&\longmapsto j,                     \nonumber
\end{align}
and we want to construct a subset of 
$\hat{L}=\{w\in h^{-1}(L) \mid \eta_0(\pi_{Y_0}(w))=\eta_1(\pi_{Y_1}(w)) \}$
in $\Alg{\HF_i}$. Observe that we cannot hope to find $\hat{L}$ itself in $\Alg{\HF_i}$ in general.
Take, for example, the context-free language $E=\{a^nb^n \mid n\ge 0\}$
and $X_0=\{a\}$, $X_1=\{b\}$, $\gamma_0(a)=1$, $\gamma_1(b)=1$. Then 
the language $\hat{E}$ would not be context-free.
However, the language $E' = \{ w\rev{g(w)}
\mid w\in\{(a,0), (a,1)\}^* \},$ where $g$ is the morphism with $(a,j)\mapsto
(b,j)$ for $j=0,1$, is context-free. Although it is only a proper subset of
$\hat{E}$, it is large enough to satisfy
$\pi_{Y_i}(E')=\pi_{Y_i}(\hat{E})=\pi_{Y_i}(h^{-1}(E))$ for $i=0,1$.  We will
see that in order to construct Parikh annotations, it suffices to use such
under-approximations of $\hat{L}$.

\newcommand{\children}[1]{\mathsf{c}(#1)}
\subparagraph{Derivation trees and matchings}
In this work, by an \emph{$X$-labeled tree}, we mean a finite ordered unranked
tree in which each node carries a label from $X\cup\{\emptyWord\}$ for an
alphabet $X$. For each node, there is a linear order on the set of its
children.  For each node $x$, we write $\children{x}\in X^*$ for the word
obtained by reading the labels of $x$'s children in this order.  Furthermore,
$\yield{x}\in X^*$ denotes the word obtained by reading leaf labels below the
node $x$ according to the linear order induced on the leaves.  Moreover, if $r$
is the root of $t$, we also write $\yield{t}$ for $\yield{r}$. The
\emph{height} of a tree is the maximal length of a path from the root to a
leaf, i.e. a tree consisting of a single node has height $0$.  A \emph{subtree}
of a tree $t$ is the tree consisting of all nodes below some node $x$ of $t$.
If $x$ is a child of $t$'s root, the subtree is a \emph{direct subtree}.

Let $G=(N,T,P,S)$ be a $\C$-grammar. A \emph{partial derivation tree (for $G$)}
is an $(N\cup T)$-labeled tree $t$ in which \begin{inparaenum}[(i)]\item each
inner node $x$ has a label $A\in N$ and there is some $A\to L$ in $P$ with
$\children{x}\in L$, and \item no $\emptyWord$-labeled node has a
sibling.\end{inparaenum}~ If, in addition, the root is labeled $S$ and every
leaf is labeled by $T\cup\{\emptyWord\}$, it is called a \emph{derivation tree
for $G$}.

Let $t$ be a tree whose leaves are $X\cup\{\emptyWord\}$-labeled. Let $L_i$
denote the set of $X_i$-labeled leaves of $t$.  An \emph{arrow collection for
$t$} is a finite set $A$ together with maps $\nu_i\colon A\to L_i$ for $i=0,1$.
Hence, $A$ can be thought of as a set of arrows pointing from $X_0$-labeled
leaves to $X_1$-labeled leaves.  We say an arrow $a\in A$ is \emph{incident} to
a leaf $\ell$ if $\nu_0(a)=\ell$ or $\nu_1(a)=\ell$.  If $\ell$ is a leaf, then
$d_A(\ell)$ denotes the number of arrows incident to $\ell$. More generally,
for a subtree $s$ of $t$, $d_A(s)$ denotes the number of arrows incident to
some leaf in $s$ and some leaf outside of $s$.  $A$ is called a
\emph{$k$-matching} if
\begin{inparaenum}[(i)]
\item each leaf labeled $x\in X_i$ has precisely $\gamma_i(x)$ incident arrows, and
\item $d_A(s)\le k$ for every subtree $s$ of $t$.
\end{inparaenum}

The following \lcnamecref{pa:matchings} applies \cref{nonterminal:extension}.
The latter implies that for nodes $x$ of a derivation tree, the balance
$\gamma_0(\pi_{X_0}(\yield{x}))-\gamma_1(\pi_{X_1}(\yield{x}))$ is bounded.
This can be used to construct $k$-matchings in a bottom-up manner.
\begin{clemma}\label{pa:matchings}
Let $X=X_0\uplus X_1$ and $\gamma_i\colon X_i^*\to\N$ for $i=0,1$ be a
morphism. Let $G$ be a reduced $\HF_i$-grammar with $\Lang{G}\subseteq X^*$
and $\gamma_0(\pi_{X_0}(w))=\gamma_1(\pi_{X_1}(w))$ for every $w\in \Lang{G}$.
Then one can compute a bound $k$ such that each derivation tree of $G$
admits a $k$-matching.  
\end{clemma}

We are now ready to construct the approximations necessary for obtaining PAIM.
\begin{cproposition}[Consistent substitution]\label{consistent:substitution}
Let $X=X_0\uplus X_1$ and $\gamma_i\colon X_i^\oplus\to\N$ for $i=0,1$ be a morphism. Let
$L\in\Alg{\HF_i}$, $L\subseteq X^*$, a language with
$\gamma_0(\pi_{X_0}(w))=\gamma_1(\pi_{X_1}(w))$ for every $w\in L$.
Furthermore, let $Y_i,h_i,\eta_i$ for $i=0,1$ and $Y,h$ be defined as in
\cref{consistent:substitution:alphabet} and
\cref{consistent:substitution:morphisms}. Moreover, let $L$ be given by a reduced grammar. Then one can construct a language
$L'\in\Alg{\HF_i}$, $L'\subseteq Y^*$, with
\begin{enumerate}[(i)]
\item\label{consistent:substitution:images} $L'\subseteq h^{-1}(L)$,
\item\label{consistent:substitution:completeness} $\pi_{Y_i}(L')=\pi_{Y_i}(h^{-1}(L))$ for $i=0,1$,
\item\label{consistent:substitution:consistency} $\eta_0(\pi_{Y_0}(w))=\eta_1(\pi_{Y_1}(w))$ for every $w\in L'$.
\end{enumerate}
\end{cproposition}
\begin{proof}
Let $G_0=(N,X,P_0,S)$ be a reduced $\HF_i$-grammar with $\Lang{G_0}=L$.
Let $G_1=(N,Y,P_1,S)$ be the grammar with  $P_1=\{A\to \hat{h}^{-1}(K) \mid A\to K\in P_0\}$,
where $\hat{h}\colon (N\cup Y)^*\to (N\cup X)^*$ is the extension of $h\colon
Y^*\to X^*$ that fixes $N$. With $L_1=\Lang{G_1}$, we clearly have
$L_1=h^{-1}(L)$. 

According to \cref{pa:matchings}, we can find a $k\in\N$ such that every
derivation tree of $G_0$ admits a $k$-matching.
With this, let $F=\{z\in\Z
\mid |z|\le k\}$, $N_2=N\times F$, and $\eta$ be the morphism
$\eta\colon (N_2\cup Y)^* \rightarrow \Z$ with
$(A,z)                    \mapsto z$ for $(A,z)\in N_2$, and
$y                       \mapsto \eta_0(\pi_{Y_0}(y))-\eta_1(\pi_{Y_1}(y))$ for $y\in Y$.
Moreover, let $g\colon (N_2\cup Y)^*\to(N\cup Y)^*$ be the morphism with
$g((A,z))=A$ for $(A,z)\in N_2$ and $g(y)=y$ for $y\in Y$.  This allows us to
define the set of productions $P_2=\{(A,z)\to g^{-1}(L)\cap\eta^{-1}(z) \mid A\to K\in P_1\}$.
Note that since $\HF_i$ is an effective Presburger closed full semi-trio, we have
effectively $g^{-1}(K)\cap\eta^{-1}(z)\in \HF_i$ for $K\in\HF_i$. Finally, let $G_2$
be the grammar $G_2=(N_2,Y,P_2,(S,0))$.  We claim that $L'=\Lang{G_2}$ has the
desired properties. Since $L'\subseteq L_1=h^{-1}(L)$,
\cref{consistent:substitution:images} is satisfied.  Furthermore, the
construction guarantees that for a production $(A,z)\to w$ in $G_2$, we have
$\eta(w)=z$.  In particular, every $w\in Y^*$ with $(S,0)\grammarsteps[G_2] w$
exhibits $\eta_0(\pi_{Y_0}(w))-\eta_1(\pi_{Y_1}(w))=\eta(w)=0$.  Thus, we have
shown \cref{consistent:substitution:consistency}.

Note that the inclusion ``$\subseteq$'' of
\cref{consistent:substitution:completeness} follows from
\cref{consistent:substitution:images}.  In order to prove 
``$\supseteq$'', we shall use $k$-matchings in $G_0$ to construct derivations
in $G_2$.  See \cref{example:matchingtrees} for an example of the following
construction of derivation trees. Let $w\in h^{-1}(L)=\Lang{G_1}$ and consider
a derivation tree $t$ for $w$ in $G_1$. Let $\bar{t}$ be the $(N\cup X)$-tree
obtained from $t$ by replacing each leaf label $y\in Y$ by $h(y)$. Then
$\bar{t}$ is a derivation tree of $G_0$ and admits a $k$-matching $\bar{A}$.
Since $\bar{t}$ and $t$ are isomorphic up to labels, we can obtain a corresponding
arrow collection $A$ in $t$ (see \cref{example:matchingtrees:a}). 

\begin{figure}
\newcommand{\examplegap}{0.5cm}
\centering
\subfloat[$t$; arrows in $A$]{
	\begin{tikzpicture}[level distance=1cm, sibling distance=0.75cm]
	\node (a) {$S$}
		child { node (u) {$(a,0)$}}
		child { node (b) {$S$}
			child { node (v) {$(a,0)$}}
			child { node (c) {$S$}
				child { node (w) {$(a,1)$}}
				child { node (d) {$S$}
					child {node (e) {$\emptyWord$}}
				}
				child { node (z) {$(b,0)$}}
			}
			child { node (y) {$(b,1)$}}
		}
		child { node (x) {$(b,0)$}};
	\draw[densely dotted, ->] (u) to [out=-60, in=-120] (x);
	\draw[densely dotted, ->] (v) to [out=-60, in=-120] (y);
	\draw[densely dotted, ->] (w) to [out=-60, in=-120] (z);
	\end{tikzpicture}
	\label{example:matchingtrees:a}
}
\hspace{\examplegap}
\subfloat[$t$; $i=1$; dashed arrow is the one in $A'$]{
	\begin{tikzpicture}[level distance=1cm, sibling distance=0.75cm]
	\node (a) {$S$}
		child { node (u) {$(a,0)$}}
		child { node (b) {$S$}
			child { node (v) {$(a,0)$}}
			child { node (c) {$S$}
				child { node (w) {$(a,1)$}}
				child { node (d) {$S$}
					child {node (e) {$\emptyWord$}}
				}
				child { node (z) {$(b,0)$}}
			}
			child { node (y) {$(b,1)$}}
		}
		child { node (x) {$(b,0)$}};
	\draw[densely dotted, ->] (u) to [out=-60, in=-120] (x);
	\draw[densely dashed, ->] (v) to [out=-60, in=-120] (y);
	\draw[densely dotted, ->] (w) to [out=-60, in=-120] (z);
	\end{tikzpicture}
	\label{example:matchingtrees:b}
}
\hspace{\examplegap}
\subfloat[$t'$]{
	\begin{tikzpicture}[level distance=1cm, sibling distance=0.75cm]
	\node (a) {$S$}
		child { node (u) {$(a,0)$}}
		child { node (b) {$S$}
			child { node (v) {$(a,1)$}}
			child { node (c) {$S$}
				child { node (w) {$(a,0)$}}
				child { node (d) {$S$}
					child {node (e) {$\emptyWord$}}
				}
				child { node (z) {$(b,0)$}}
			}
			child { node (y) {$(b,1)$}}
		}
		child { node (x) {$(b,0)$}};

	\end{tikzpicture}
	\label{example:matchingtrees:c}
}
\hspace{\examplegap}
\subfloat[$t''$]{
	\begin{tikzpicture}[level distance=1cm, sibling distance=1cm]
	\node (a) {$(S,0)$}
		child { node (u) {$(a,0)$}}
		child { node (b) {$(S,0)$}
			child { node (v) {$(a,1)$}}
			child { node (c) {$(S,0)$}
				child { node (w) {$(a,0)$}}
				child { node (d) {$(S,0)$}
					child {node (e) {$\emptyWord$}}
				}
				child { node (z) {$(b,0)$}}
			}
			child { node (y) {$(b,1)$}}
		}
		child { node (x) {$(b,0)$}};
	\end{tikzpicture}
	\label{example:matchingtrees:d}
}
\caption{Derivation trees in the proof of \cref{consistent:substitution} for
the context-free grammar $G$ with productions $S\to aSb$, $S\to\emptyWord$ and
$X_0=\{a\}$, $X_1=\{b\}$, $\gamma_0(a)=\gamma_1(b)=1$.}
\label{example:matchingtrees}
\end{figure}
Let $L_i$ denote the set of $Y_i$-labeled leaves of $t$ for $i=0,1$.  Now fix
$i\in\{0,1\}$. We choose a subset $A'\subseteq A$ as follows. Since $\bar{A}$
is a $k$-matching, each leaf $\ell\in L_{i}$ of $t$ has precisely
$\gamma_{i}(h(\lambda(\ell)))\ge\eta_{i}(\lambda(\ell))$ incident arrows in
$A$.  For each such $\ell\in L_{i}$, we include some arbitrary choice of
$\eta_{i}(\lambda(\ell))$ arrows in $A'$ (see
\cref{example:matchingtrees:b}).  The tree $t'$ is obtained from $t$ by
changing the label of each leaf $\ell\in L_{1-i}$ from $(x,j)$ to $(x,j')$,
where $j'$ is the number of arrows in $A'$ incident to $\ell$ (see
\cref{example:matchingtrees:c}). Note that since we only change labels of
leaves in $L_{1-i}$, we have $\pi_{Y_i}(\yield{t'})=\pi_{Y_i}(\yield{t})=\pi_{Y_i}(w)$.

For every subtree $s$ of $t'$, we define
$\beta(s)=\eta_0(\pi_{Y_0}(\yield{s}))-\eta_1(\pi_{Y_1}(\yield{s}))$.  By construction of $A'$, each
leaf $\ell\in L_j$ has precisely $\eta_j(\lambda(\ell))$ incident arrows in
$A'$ for $j=0,1$. Therefore,
\begin{equation}\beta(s)=\sum_{\ell\in L_0\cap s} d_{A'}(\ell)-\sum_{\ell\in L_1\cap s} d_{A'}(\ell). \label{pa:eq:eta:degrees}\end{equation}
The absolute value of the right hand side of this equation is at
most $d_{A'}(s)$ and hence 
\begin{equation} |\eta_0(\pi_{Y_0}(\yield{s}))-\eta_1(\pi_{Y_1}(\yield{s}))|=|\beta(s)|\le d_{A'}(s)\le d_A(s)\le k \label{pa:eq:beta:bound}\end{equation}
since $\bar{A}$ is a $k$-matching. In the case $s=t'$, 
\cref{pa:eq:eta:degrees} also tells us that
\begin{equation} \eta_0(\pi_{Y_0}(\yield{t'}))-\eta_1(\pi_{Y_1}(\yield{t'}))=\sum_{\ell\in L_0} d_{A'}(\ell)-\sum_{\ell\in L_1} d_{A'}(\ell)=0. \label{pa:eq:eta:zero}\end{equation}

Let $t''$ be the tree obtained from $t'$ as follows: For each $N$-labeled node $x$ of
$t'$, we replace the label $B$ of $x$ with $(B,\beta(s))$, where $s$ is the subtree below $x$ (see
\cref{example:matchingtrees:d}).  By \cref{pa:eq:beta:bound}, this is a symbol
in $N_2$. The root node of $t''$ has label $(S,0)$ by
\cref{pa:eq:eta:zero}. Furthermore, it follows by an induction on the hight of
subtrees that if $(B,z)$ is the label of a node $x$, then $z=\eta(\children{x})$. Hence, the tree $t''$
is a derivation tree of $G_2$. This means $\pi_{Y_i}(w)=\pi_{Y_i}(\yield{t'})=\pi_{Y_i}(\yield{t''})\in L(G_2)=L'$,
completing the proof of \cref{consistent:substitution:completeness}.
\end{proof}

\Cref{consistent:substitution} now allows us to construct PAIM for languages
$\sigma(L)$, where $\sigma$ is a letter substitution. 
The essential idea is to
use a PAIM $(K,C,P,(P_c)_{c\in C},\varphi,\marker)$ for $L$ and then apply
\cref{consistent:substitution} to $K$ with $X_0=Z\cup\{\marker\}$ and
$X_1=C\cup P$. One can clearly assume that a single
letter $a$ from $Z$ is replaced by $\{a,b\}\subseteq Z'$.  We can therefore
choose $\gamma_0(w)$ to be the number of $a$'s in $w$ and $\gamma_1(w)$ to be
the number of $a$'s represented by symbols in $C\cup P$. Then the counting
property of $K$ entails $\gamma_0(w)=\gamma_1(w)$ for $w\in K$ and thus
applicability of \cref{consistent:substitution}.
\Cref{consistent:substitution:completeness} then yields the projection property
for $i=0$ and the commutative projection property for $i=1$ and
\cref{consistent:substitution:consistency} yields the counting property for the
new PAIM.
\begin{clemma}[Letter substitution]\label{pa:lettersubstitution}
Let $\sigma\colon Z\to\Powerset{Z'}$ be a letter substitution. Given $i\in\N$
and a PAIM for $L\in\HG_i$ in $\HG_i$, one can construct a PAIM in
$\HG_i$ for $\sigma(L)$.
\end{clemma}

The basic idea for the case of general substitutions is to replace each $x$ by a PAIM for
$\sigma(x)$.
Here, 
\cref{pa:lettersubstitution} allows us to
assume that the PAIM for each $\sigma(x)$ is linear. However, we have to make
sure that the number of occurrences of $\marker$ remains bounded.
\begin{clemma}[Substitutions]\label{pa:substitution}
Let $L\subseteq X^*$ in $\HG_i$ and $\sigma$ be a $\HG_i$-substitution. Given a
PAIM in $\HG_i$ for $L$ and for each $\sigma(x)$, $x\in X$, one can construct a
PAIM for $\sigma(L)$ in $\HG_i$.
\end{clemma}

The next step is to construct PAIM for languages $\Lang{G}$, where $G$ has just
one nonterminal $S$ and PAIM are given for the right-hand-sides. 
Here, it
suffices to obtain a PAIM for $\SententialForms{G}$ in the case that $S$ occurs
in every word on the right hand side: Then $\Lang{G}$ can be obtained from
$\SententialForms{G}$ using a substitution.  Applying $S\to R$ then means that
for some $w\in R$, $\Parikh{w}-S$ is added to the Parikh image of the
sentential form. Therefore, computing a PAIM for $\SententialForms{G}$ is akin
to computing a semilinear representation for $S^\oplus$, where $S$ is
semilinear.
\begin{clemma}[One nonterminal]\label{pa:onenonterminal}
Let $G$ be a $\HG_i$-grammar with one nonterminal.  Furthermore,
suppose PAIM in $\HG_i$ are given for the right-hand-sides in $G$. Then
we can construct a PAIM for $\Lang{G}$ in $\HG_i$.
\end{clemma}

Using \cref{pa:substitution,pa:onenonterminal}, we can now construct PAIM
recursively with respect to the number of nonterminals in $G$.
\begin{clemma}[PAIM for algebraic extensions]\label{pa:alg}
Given $i\in\N$ and an $\HF_i$-grammar $G$, along with a PAIM in $\HF_i$ for
each right hand side, one can construct a PAIM for $\Lang{G}$ in $\HG_i$.
\end{clemma}

The last step is to compute PAIM for languages in $\HomSLI{\HG_i}$.  Then, \cref{pa:parikhannotations} follows.
\begin{clemma}[PAIM for semilinear intersections]\label{pa:sli}
Given $i\in\N$, a language $L\subseteq X^*$ in $\HG_i$, a semilinear
$S\subseteq X^\oplus$, and a morphism $h\colon X^*\to Y^*$, along with a
PAIM in $\HG_i$ for $L$, one can construct a PAIM for $h(L\cap\ParikhInv{S})$
in $\HomSLI{\HG_i}$.
\end{clemma}

\section{Computing downward closures}
\label{sec:dclosure}
The procedure for computing downward closures works recursively with respect to
the hierarchy $\HF_0\subseteq\HG_0\subseteq\cdots$.  For languages in
$\HG_i=\Alg{\HF_i}$, we use an idea by van~Leeuwen~\cite{vanLeeuwen1978}, who
proved that downward closures are computable for $\Alg{\C}$ if and only if this
is the case for $\C$.
This means we can compute downward closures for $\HG_i$ if we can compute them for $\HF_i$.
For the latter, we use \cref{dc:overapprox}, which is based on the following idea.
Using a PAIM for $L$ in $\HG_i$, 
one constructs a language $L'\supseteq L\cap\ParikhInv{S}$ in which every
word admits insertions that yield a word in $L\cap\ParikhInv{S}$, meaning that $\Dclosure{L'}=\Dclosure{(L\cap \ParikhInv{S})}$. Here, 
$L'$ is obtained from the PAIM using a rational transduction, which implies $L'\in\HG_i$.
\begin{clemma}\label{dc:overapprox}
Given $i\in\N$, a language $L\subseteq X^*$ in $\HG_i$, and a semilinear
$S\subseteq X^\oplus$, one can compute a language $L'\in\HG_i$ with
$\Dclosure{L'}=\Dclosure{(L\cap\ParikhInv{S})}$.
\end{clemma}

\begin{ctheorem}\label{dc:dclosure}
Given a language $L$ in $\HF$, one can compute a finite automaton for $\Dclosure{L}$.
\end{ctheorem}
\begin{proof}
We perform the computation recursively with respect to the level of the
hierarchy $\HF_0\subseteq\HG_0\subseteq\HF_1\subseteq\HG_1\subseteq\cdots$.
\begin{itemize}
\item If $L\in\HF_0$, then $L$ is finite and we can clearly compute $\Dclosure{L}$.
\item If $L\in\HF_i$ with $i\ge 1$, then $L=h(L'\cap\ParikhInv{S})$ for some
$L'\subseteq X^*$ in $\HG_{i-1}$, a semilinear $S\subseteq X^\oplus$, and a
morphism $h$.  Since $\Dclosure{h(M)}=\Dclosure{h(\Dclosure{M})}$ for any $M\subseteq X^*$, it suffices to
describe how to compute 
$\Dclosure{(L'\cap\ParikhInv{S})}$. Using \cref{dc:overapprox}, we construct a
language $L''\in\HG_{i-1}$ with
$\Dclosure{L''}=\Dclosure{(L'\cap\ParikhInv{S})}$ and then recursively compute
$\Dclosure{L''}$.
\item If $L\in\HG_i$, then $L$ is given by an $\HF_i$-grammar $G$. Using
recursion, we compute the downward closure of each right-hand-side of $G$. We
obtain a new $\Reg$-grammar $G'$ by replacing each right-hand-side in $G$ with
its downward closure.  Then $\Dclosure{\Lang{G'}}=\Dclosure{L}$.  Since we can
construct a context-free grammar for $\Lang{G'}$, we can compute
$\Dclosure{\Lang{G'}}$ using the available algorithms by
van~Leeuwen~\cite{vanLeeuwen1974} or Courcelle~\cite{Courcelle1991}.
\end{itemize}
\end{proof}

\section{Strictness of the hierarchy}
\label{sec:strictness}
In this section, we present another application of Parikh annotations.  Using
PAIM, one can show that the inclusions
$\HF_0\subseteq\HG_0\subseteq\HF_1\subseteq\HG_1\subseteq\cdots$ in the
hierarchy are, in fact, all strict. 
It is of course easy to see that
$\HF_0\subsetneq\HG_0\subsetneq\HF_1$, since $\HF_0$ contains only finite sets
and $\HF_1$ contains, for example, $\{a^nb^nc^n\mid n\ge 0\}$. 
In order to
prove strictness at higher levels, we present two transformations: The first
turns a language from $\HF_i\setminus\HG_{i-1}$ into one in
$\HG_{i}\setminus\HF_i$ (\cref{strictness:sli}) and the second turns one from
$\HG_i\setminus\HF_{i}$ into one in $\HF_{i+1}\setminus\HG_i$
(\cref{strictness:shuffle}).

The essential idea of the next \lcnamecref{strictness:sli} is as follows. For
the sake of simplicity, assume $(L\#)^*=L'\cap\ParikhInv{S}$ for $L'\in\C$,
$L'\subseteq (X\cup \{\#\})^*$. Consider a PAIM $(K',C,P,(P_c)_{c\in
C},\varphi,\marker)$ for $L'$ in $\C$. Using a rational transduction, we obtain
from $K'$ a language $\hat{L}\subseteq (X\cup\{\#,\marker\})^*$ in $\C$ such
that every member of $\hat{L}$ admits an insertion at $\marker$ that yields a word
from $(L\#)^*=L'\cap\ParikhInv{S}$. Using rational transductions again, we can
then pick all words that appear between two $\#$ in some member of $\hat{L}$ and contain no
$\marker$.  Since there is a bound on the number of $\marker$ in $K'$ (and
hence in $\hat{L}$), every word from $L$ has to occur in this way. On the other
hand, since inserting at $\marker$ yields a word in $(L\#)^*$, every such word without
$\marker$ must be in $L$.
\begin{cproposition}\label{strictness:sli}
Let $\C$ be a full trio such that every language in $\C$ has a PAIM in $\C$.
Moreover, let $X$ be an alphabet with $\#\notin X$.  If
$(L\#)^*\in\HomSLI{\C}$ for $L\subseteq X^*$, then $L\in\C$.
\end{cproposition}

In order to prove \cref{strictness:shuffle}, we need a new concept.  A bursting
grammar is one in which essentially (meaning: aside from a subsequent
replacement by terminal words of bounded length) the whole word is generated in
a single application of a production.
\begin{cdefinition}
Let $\C$ be a language class and $k\in\N$. A $\C$-grammar $G$ is called
\emph{$k$-bursting} if for every derivation tree $t$ for $G$ and every node
$x$ of $t$ we have: $|\yield{x}|>k$ implies $\yield{x}=\yield{t}$. A grammar is
said to be \emph{bursting} if it is $k$-bursting for some $k\in\N$.
\end{cdefinition}
\begin{clemma}\label{strictness:bursting}
If $\C$ is a union closed full semi-trio and $G$ a bursting $\C$-grammar, then
$\Lang{G}\in\C$.
\end{clemma}

The essential idea for \cref{strictness:shuffle} is the
following.  We construct a $\C$-grammar $G'$ for $L$ by removing from a
$\C$-grammar $G$ for $M=(L\shuffle\{a^nb^nc^n\mid n\ge 0\})\cap a^*(bX)^*c^*$
all terminals $a,b,c$.  Using \cref{nonterminal:extension}, one can then show that $G'$ is bursting.
\begin{cproposition}\label{strictness:shuffle}
Let $\C$ be a union closed full semi-trio and let $a,b,c\notin X$ and
$L\subseteq X^*$.  If $L\shuffle\{a^nb^nc^n\mid n\ge 0\} \in \Alg{\C},$ then
$L\in\C$.
\end{cproposition}

We can now show that the hierarchy $\HF_0\subseteq\HG_0\subseteq\HF_1\subseteq\HG_1\subseteq\cdots$ is strict.

\begin{ctheorem}\label{strictness:hierarchy}
For $i\in\N$, define the alphabets $X_0=\emptyset$,
$Y_i=X_i\cup\{\#_i\}$, $X_{i+1}=Y_i\cup\{a_{i+1},b_{i+1},c_{i+1}\}$.  Moreover,
define $U_i\subseteq X_i^*$ and $V_i\subseteq Y_i^*$ as $U_0=\{\emptyWord\}$,
$V_i=(U_i\#_i)^*$, and $U_{i+1}=V_i\shuffle \{a_{i+1}^nb_{i+1}^nc_{i+1}^n \mid
n\ge 0\}$ for $i\ge 0$. Then $V_i\in \HG_i\setminus\HF_i$ and $U_{i+1}\in
\HF_{i+1}\setminus\HG_i$.
\end{ctheorem}

\printbibliography

\appendix

\section{Proof of \cref{algbb}}

In order to prove \cref{algbb}, we define the relevant notions in detail.

Let $A$ be a (not necessarily finite) set of symbols and $R\subseteq A^*\times
A^*$.  The pair $(A,R)$ is called a \emph{(monoid) presentation}. The smallest
congruence of $A^*$ containing $R$ is denoted by $\congruence_R$ and we will
write $[w]_R$ for the congruence class of $w\in A^*$.  The \emph{monoid
presented by $(A,R)$} is defined as $A^*/\mathord{\congruence_R}$. For the
monoid presented by $(A,R)$, we also write $\langle A \mid R\rangle$, where $R$
is denoted by equations instead of pairs.

Note that since we did not impose a finiteness restriction on $A$, every monoid
has a presentation. Furthermore, for monoids $M_1$, $M_2$ we can find
presentations $(A_1,R_1)$ and $(A_2,R_2)$ such that $A_1\cap A_2=\emptyset$.
We define the \emph{free product} $M_1*M_2$ to be presented by $(A_1\cup A_2,
R_1\cup R_2)$. Note that $M_1*M_2$ is well-defined up to isomorphism. By way of
the injective morphisms $[w]_{R_i}\mapsto [w]_{R_1\cup R_2}$, $w\in A_i^*$ for
$i=1,2$, we will regard $M_1$ and $M_2$ as subsets of $M_1*M_2$.  It is a
well-known property of free products that if $\varphi_i\colon M_i\to N$ is a
morphism for $i=1,2$, then there is a unique morphism $\varphi\colon M_1*M_2\to
N$ with $\varphi|_{M_i}=\varphi_i$ for $i=1,2$.  Furthermore, if
$u_0v_1u_1\cdots v_nu_n=1$ for $u_0,\ldots,u_n\in M_1$ and $v_1,\ldots,v_n\in
M_2$ (or vice versa), then $u_j=1$ or $v_j=1$ for some $0\le j\le n$.
Moreover, we write $M^{(n)}$ for the $n$-fold free product $M*\cdots*M$.

One of the directions of the equality $\VA{\B*\B*M}=\Alg{\VA{M}}$ follows from
previous work.  In \cite{Zetzsche2013a} (and, for a more general
product construction, in \cite{BuckheisterZetzsche2013a}), the following was
shown.
\begin{ctheorem}[\cite{Zetzsche2013a,BuckheisterZetzsche2013a}]\label{basics:freeproduct}
Let $M_0$ and $M_1$ be monoids. Then $\VA{M_0 * M_1}\subseteq\Alg{\VA{M_0}\cup\VA{M_1}}$.
\end{ctheorem}

Let $M$ and $N$ be monoids. In the following, we write $M\hookrightarrow N$ if
there is a morphism $\varphi\colon M\to N$ such that $\varphi^{-1}(1)=\{1\}$.
Clearly, if $M\hookrightarrow N$, then $\VA{M}\subseteq\VA{N}$: Replacing in a
valence automaton over $M$ all elements $m\in M$ with $\varphi(m)$ yields a
valence automaton over $N$ that accepts the same language.
\begin{clemma}
If $M\hookrightarrow M'$ and $N\hookrightarrow N'$, then $M*N\hookrightarrow M'*N'$.
\end{clemma}
\begin{proof}
Let $\varphi\colon M\to M'$ and $\psi\colon N\to N'$. Then the morphism
$\kappa\colon M*N\to M'*N'$ with $\kappa|_M=\varphi$ and $\kappa|_N=\psi$
clearly satisfies $\kappa^{-1}(1)=1$.
\end{proof}

We will use the notation $\Rclass_1(M)=\{a\in M\mid \exists b\in M\colon ab=1\}$.
\begin{clemma}\label{bpowers}
Let $M$ be a monoid with $\Rclass_1(M)\ne\{1\}$. Then
$\B^{(n)}*M\hookrightarrow \B*M$ for every $n\ge
1$.  In particular, $\VA{\B*M}=\VA{\B^{(n)}*M}$ for every $n\ge 1$.
\end{clemma}
\begin{proof}
If $\B^{(n)}*M\hookrightarrow \B*M$ and $\B*\B*M\hookrightarrow \B*M$,
then
\[ \B^{(n+1)}*M\cong \B*(\B^{(n)}*M)\hookrightarrow \B*(\B*M)\hookrightarrow \B*M. \]
Therefore, it suffices to prove $\B*\B*M\hookrightarrow \B*M$.

Let $\B_s=\langle s,\bar{s}\mid s\bar{s}=1\rangle$ for $s\in\{p,q,r\}$. We show
$\B_p*\B_q*M\hookrightarrow \B_r*M$.  Suppose $M$ is presented by $(X,R)$.  We
regard the monoids $\B_p*\B_q*M$ and $\B_r*M$ as embedded into
$\B_p*\B_q*\B_r*M$, which by definition of the free product, has a presentation
$(Y,S)$, where $Y=\{p,\bar{p},q,\bar{q},r,\bar{r}\}\cup X$ and $S$ consists of
$R$ and the equations $s\bar{s}=1$ for $s\in \{p,q,r\}$.  For $w\in Y^*$, we
write $[w]$ for the congruence class generated by $S$.  Since
$\Rclass_1(M)\ne\{1\}$, we find $u,v\in X^*$ with $[uv]=1$ and $[u]\ne 1$.  and
let $\varphi\colon (\{p,\bar{p},q,\bar{q}\}\cup X)^*\to(\{r,\bar{r}\}\cup X)^*$
be the morphism with $\varphi(x)=x$ for $x\in X$ and
\begin{align*}
p&\mapsto rr, & \bar{p}&\mapsto \bar{r}\bar{r}, \\ 
q&\mapsto rur, & \bar{q}&\mapsto \bar{r}v\bar{r}.
\end{align*}
We show by induction on $|w|$ that $[\varphi(w)]=1$ implies $[w]=1$. Since this
is trivial for $w=\emptyWord$, we assume $|w|\ge 1$.  Now suppose
$[\varphi(w)]=[\emptyWord]$ for some $w\in(\{p,\bar{p},q,\bar{q}\}\cup X)^*$. If $w\in X^*$,
then $[\varphi(w)]=[w]$ and hence $[w]=1$.  Otherwise, we have
$\varphi(w)=xry\bar{r}z$ for some $y\in X^*$ with $[y]=1$ and $[xz]=1$. This
means $w=fsy\overline{s'}g$ for $s,s'\in\{p,q\}$ with $\varphi(fs)=xr$ and
$\varphi(\overline{s'}g)=\bar{r}z$. If $s\ne s'$, then $s=p$ and $s'=q$; or
$s=q$ and $s'=p$.  In the former case
\[ [\varphi(w)]=[\varphi(f)~rr~y~\bar{r}v\bar{r}~\varphi(g)]=[\varphi(f)rv\bar{r}\varphi(g)]\ne 1 \]
since $[v]\ne 1$ and in the latter
\[ [\varphi(w)]=[\varphi(f)~rur~y~\bar{r}\bar{r}~\varphi(g)]=[\varphi(f)ru\bar{r}\varphi(g)]\ne 1 \]
since $[u]\ne 1$. Hence $s=s'$. This means $1=[w]=[fsy\bar{s}g]=[fg]$ and
$1=[\varphi(w)]=[\varphi(fg)]$ and since $|fg|<|w|$, induction yields
$[w]=[fg]=1$.

Hence, we have shown that $[\varphi(w)]=1$ implies $[w]=1$. Since, on
the other hand, $[u]=[v]$ implies $[\varphi(u)]=[\varphi(v)]$ for all $u,v\in
(\{p,\bar{p},q,\bar{q}\}\cup X)^*$, we can lift $\varphi$ to a morphism
witnessing $\B_p*\B_q*M\hookrightarrow \B_r*M$.
\end{proof}

\begin{proof}[Proof of \cref{algbb}]
It suffices to prove the first statement: If $\Rclass_1(M)\ne\{1\}$, then by
\cref{bpowers}, $\VA{\B*M}=\VA{\B*\B*M}$.
Since $\VA{\B}\subseteq\CF$, \cref{basics:freeproduct} yields
\[ \VA{\B*N}\subseteq\Alg{\VA{\B}\cup\VA{N}}\subseteq\Alg{\VA{N}} \]
for every monoid $N$. Therefore,
\[ \VA{\B*\B*M}\subseteq\Alg{\VA{\B*M}}\subseteq\Alg{\Alg{\VA{M}}}=\Alg{\VA{M}}. \]
It remains to be shown that $\Alg{\VA{M}}\subseteq\VA{\B*\B*M}$.

Let $G=(N,T,P,S)$ be a reduced $\VA{M}$-grammar and let $X=N\cup T$. Since
$\VA{M}$ is closed under union, we may assume that for each $B\in N$, there is
exactly one production $B\to L_B$ in $P$.  For each $B\in N$, let $A_B=(Q_B,X,M,E_B,q^B_0,F_B)$ by a
valence automaton over $M$ with $\Lang{A_B}=L_B$.  We may clearly assume that
$Q_B\cap Q_C=\emptyset$ for $B\ne C$ and that for each $(p,w,m,q)\in E_B$,
we have $|w|\le 1$.

\newcommand{\bleft}{\lfloor}
\newcommand{\bright}{\rfloor}
In order to simplify the correctness proof, we modify $G$.  Let $\bleft$ and
$\bright$ be new symbols and let $G'$ be the grammar
$G'=(N,T\cup\{\bleft,\bright\},P',S)$, where $P'$ consists of the productions
$B\to \bleft L\bright$ for $B\to L\in P$.
Moreover, let
\[ K=\{v\in (N\cup T\cup\{\bleft,\bright\})^* \mid u \grammarsteps[G'] v,~u\in L_S\}. \]
Then $\Lang{G}=\pi_T(K\cap (T\cup\{\bleft,\bright\})^*)$ and it suffices to
show $K\in\VA{\B*\B*M}$.

Let $Q=\bigcup_{B\in N} Q_B$. For each $q\in Q$, let $\B_q=\langle
q,\bar{q}\mid q\bar{q}=1\rangle$ be an isomorphic copy of $\B$.  Let
$M'=\B_{q_1}*\cdots*\B_{q_n}*M$, where $Q=\{q_1,\ldots,q_n\}$.  We shall prove
$K\in\VA{M'}$, which implies $K\in\VA{\B*\B*M}$ by \cref{bpowers} since $\Rclass_1(\B*M)\ne\{1\}$.

Let $E=\bigcup_{B\in N} E_B$, $F=\bigcup_{B\in N} F_B$.  The new set $E'$
consists of the following transitions:
\begin{align}
&(p,x,m,q)              && \text{for $(p,x,m,q)\in E$,} \label{basics:algbb:t:old} \\
&(p,\bleft,mq,q^B_0)    && \text{for $(p,B,m,q)\in E$, $B\in N$,} \label{basics:algbb:t:open} \\
&(p,\bright,\bar{q},q)  && \text{for $p\in F$, $q\in Q$.} \label{basics:algbb:t:close}
\end{align}
We claim that with $A'=(Q, N\cup T\cup\{\bleft,\bright\},M',E',q_0^S,F)$, we have
$\Lang{A'}=K$.

Let $v\in K$, where $u\grammarstepsn[G']{n} v$ for some $u\in L_S$. We show
$v\in\Lang{A'}$ by induction on $n$. For $n=0$, we have $v\in L_S$ and can use
transitions of type \labelcref{basics:algbb:t:old} inherited from $A_S$ to
accept $v$. If $n\ge 1$, let $u\grammarstepsn[G']{n-1} v'\grammarstep[G'] v$.
Then $v'\in \Lang{A'}$ and $v'=xBy$, $v=x\bleft w\bright y$ for some $B\in N$,
$w\in L_B$. The run for $v'$ uses a transition $(p,B,m,q)\in E$. Instead of
using this transition, we can use $(p,\bleft,mq,q_0^B)$, then execute the
\labelcref{basics:algbb:t:old}-type transitions for $w\in L_B$, and finally use
$(f,\bright,\bar{q},q)$, where $f$ is the final state in the run for $w$.  This
has the effect of reading $\bleft w\bright$ from the input and multiplying
$mq1\bar{q}=m$ to the storage monoid.  Hence, the new run is valid and accepts
$v$. Hence, $v\in \Lang{A'}$. This proves $K\subseteq\Lang{A'}$.

In order to show $\Lang{A'}\subseteq K$, consider the morphisms
$\varphi\colon (T\cup\{\bleft,\bright\})^*\to \B$, $\psi\colon M'\to\B$ with
$\varphi(x)=1$ for $x\in T$, $\varphi(\bleft)=a$, $\varphi(\bright)=\bar{a}$,
$\psi(q)=a$ for $q\in Q$, $\psi(\bar{q})=\bar{a}$, and $\psi(m)=1$ for $m\in
M$.  The transitions of $A'$ are constructed such that
$(p,\emptyWord,1)\autsteps[A'] (q,w,m)$ implies $\varphi(w)=\psi(m)$. In
particular, if $v\in \Lang{A'}$, then $\pi_{\{\bleft,\bright\}}(v)$ is a semi-Dyck
word with respect to $\bleft$ and $\bright$.

Let $v\in\Lang{A'}$ and let $n=|w|_{\bleft}$. We show $v\in K$ by induction on
$n$.  If $n=0$, then the run for $v$ only used transitions of type
\labelcref{basics:algbb:t:old} and hence $v\in L_S$. If $n\ge 1$, since
$\pi_{\{\bleft,\bright\}}(v)$ is a semi-Dyck word, we can write $v=x\bleft
w\bright y$ for some $w\in (N\cup T)^*$. Since $\bleft$ and $\bright$ can only
be produced by transitions of the form \labelcref{basics:algbb:t:open} and
\labelcref{basics:algbb:t:close}, respectively, the run for $v$ has to be of
the form
\begin{align*}
 (q_0^S,\emptyWord,1)&\autsteps[A'](p,x,r) \\
                     &\autstep[A'](q_0^B,x\bleft,rmq) \\
                     &\autsteps[A'](f,x\bleft w,rmqs) \\
                     &\autstep[A'](q',x\bleft w\bright,rmqs\overline{q'}) \\
                     &\autsteps[A'] (f',x\bleft w\bright y,rmqs\overline{q'}t)
\end{align*}
for some $p,q,q'\in Q$, $B\in N$, $(p,B,m,q)\in E$, $f,f'\in F$, $r,t\in M'$,
and $s\in M$ and with $rmqs\overline{q'}t=1$. This last condition implies $s=1$
and $q=q'$, which in turn entails $rmt=1$. This also means
$(p,B,m,q')=(p,B,m,q)\in E$ and $(q_0^B,\emptyWord,1)\autsteps[A']
(f,w,s)=(f,w,1)$ and hence $w\in L_B$.  Using the transition $(p,B,m,q')\in E$, we
have
\begin{align*}
(q_0^S,\emptyWord,1) &\autsteps[A'] (p,x,r) \\
                     &\autstep[A'] (q',xB,rm) \\
                     &\autsteps[A'] (f',xBy,rmt).
\end{align*}
Hence $xBy\in\Lang{A'}$ and $|xBy|_{\bleft}<|v|_{\bleft}$. Thus, induction
yields $xBy\in K$ and since $xBy\grammarstep[G'] x\bleft w\bright y$, we have $v=x\bleft w\bright y\in K$.  This
establishes $\Lang{A'}=K$.
\end{proof}

\section{Proof of \cref{basics:sli:zpowers}}
\begin{proof}
We start with the inclusion ``$\subseteq$''. Since the right-hand side is
closed under morphisms and union, it suffices to show that for each
$L\in\VA{M}$, $L\subseteq X^*$, and semilinear $S\subseteq X^\oplus$, we have
$L\cap\ParikhInv{S}\in\VA{M\times\Z^n}$ for some $n\ge 0$. Let $n=|X|$ and pick a
linear order on $X$. This induces an embedding $X^\oplus\to\Z^n$,
by way of which we consider $X^\oplus$ as a subset of $\Z^n$.

Suppose $L=\Lang{A}$ for a valence automaton $A$ over $M$. The new valence
automaton $A'$ over $M\times\Z^n$  simulates $A$ and, if $w$ is the input read
by $A$, adds $\Parikh{w}$ to the $\Z^n$ component of the storage monoid.  When
$A$ reaches a final state, $A'$ nondeterministically changes to a new state
$q_1$, in which it nondeterministically subtracts an element of $S$ from the
$\Z^n$ component.  Afterwards, $A'$ switches to another new state $q_2$, which
is the only accepting state in $A'$. Clearly, $A'$ accepts a word $w$ if and
only if $w\in \Lang{A}$ and $\Parikh{w}\in S$, hence
$\Lang{A'}=\Lang{A}\cap\ParikhInv{S}$. This proves ``$\subseteq$''.

Suppose $L=\Lang{A}$ for some valence automaton $A=(Q,X,M\times\Z^n,E,q_0,F)$.
We construct a valence automaton $A'$ over $M$ as follows. The input alphabet
$X'$ of $A'$ consists of all those $(w,\mu)\in X^*\times\Z^n$ for which there
is an edge $(p,w,(m,\mu),q)\in E$ for some $p,q\in Q$, $m\in M$. $A'$ has edges
\[ E'=\{ (p, (w,\mu), m, q) \mid (p, w, (m,\mu), q)\in E \}. \] In other words,
whenever $A$ reads $w$ and adds $(m,\mu)\in M\times\Z^n$ to its storage monoid,
$A'$ adds $m$ and reads $(w,\mu)$ from the input.  Let $\psi\colon
X'^\oplus\to\Z^n$ be the morphism that projects the symbols in $X'$ to the
right component and let $h\colon X'^*\to X^*$ be the morphism that projects the
symbols in $X'$ to the left component. Note that the set
$S=\psi^{-1}(0)\subseteq X'^\oplus$ is Presburger definable and hence
effectively semilinear. We clearly have
$\Lang{A}=h(\Lang{A'}\cap\ParikhInv{S})\in\HomSLI{\VA{M}}$.  This proves
``$\supseteq$''. Clearly, all constructions in the proof can be
carried out effectively.
\end{proof}

\section{Proof of \cref{hierarchy:closure}}
\begin{cproposition}\label{basics:alg:semiafl}
Let $\C$ be an effective full semi-trio. Then $\Alg{\C}$ is an effective full semi-AFL.
\end{cproposition}
\begin{proof}
Since $\Alg{\C}$ is clearly effectively closed under union, we only prove
effective closure under rational transductions.

Let $G=(N,T,P,S)$ be a $\C$-grammar and let $U\subseteq X^*\times T^*$ be a
rational transduction.  Since we can easily construct a $\C$-grammar for
$a\Lang{G}$ (just add a production $S'\to \{aS\}$) and the rational
transduction $(\emptyWord, a)U=\{ (v,au) \mid (v,u)\in U\}$, we may assume that
$\Lang{G}\subseteq T^+$.

Let $U$ be given by the automaton $A=(Q,X^*\times T^*,E,q_0,F)$. We may assume that
\[ E \subseteq Q\times ((X\times\{\emptyWord\})\cup (\{\emptyWord\}\times T))\times Q \]
and $F=\{f\}$.
We regard $Z=Q\times T\times Q$ and $N'=Q\times N\times Q$ as alphabets. For
each $p,q\in Q$, let $U_{p,q}\subseteq N'\times (N\cup T)^*$ be the transduction
such that for $w=w_1\cdots
w_n$, $w_1,\ldots,w_n\in N\cup T$, $n\ge 1$, the set $U_{p,q}(w)$ consists of all words
\[ (p,w_1,q_1)(q_1,w_2,q_2)\cdots (q_{n-1},w_n,q)  \]
with $q_1,\ldots,q_{n-1}\in Q$. Moreover, let
$U_{p,q}(\emptyWord)=\{\emptyWord\}$ if $p=q$ and $U_{p,q}(\emptyWord)=\emptyset$ if $p\ne q$.
Observe that $U_{p,q}$ is locally finite.
The new grammar $G'=(N',Z,P',(q_0,S,f))$ has productions $(p,B,q)\to
U_{p,q}(L)$ for each $p,q\in Q$ and $B\to L\in P$.
Let $\sigma\colon Z^*\to\Powerset{X^*}$ be the regular substitution defined by
\[ \sigma((p,x,q)) = \{ w\in X^* \mid (p,(\emptyWord,\emptyWord))\autsteps[A] (q, (w, x)) \}. \]
We claim that $U(\Lang{G})=\sigma(\Lang{G'})$. First, it can be shown by
inducion on the number of derivation steps that
$\SententialForms{G'}=U_{q_0,f}(\SententialForms{G})$. This implies
$\Lang{G'}=U_{q_0,f}(\Lang{G})$.  Since for every language $K\subseteq T^+$, we
have $\sigma(U_{q_0,f}(K))=UK$, we may conclude
$\sigma(\Lang{G'})=U(\Lang{G})$.

$\Alg{\C}$ is clearly effectively closed under $\Alg{\C}$-substitutions.  Since
$\C$ contains the finite languages, this means $\Alg{\C}$ is closed under
$\Reg$-substitutions. Hence, we can construct a $\C$-grammar for
$U(\Lang{G})=\sigma(\Lang{G'})$.
\end{proof}

\begin{cproposition}\label{basics:sli:presburger:trio}
Let $\C$ be an effective full semi-AFL. Then $\HomSLI{\C}$ is an effective
Presburger closed full trio. In particular,
$\HomSLI{\HomSLI{\C}}=\HomSLI{\C}$.
\end{cproposition}
\begin{proof}
Let $L\in\C$, $L\subseteq X^*$, $S\subseteq X^\oplus$ semilinear, and $h\colon
X^*\to Y^*$ be a morphism.  If $T\subseteq Z^*\times Y^*$ is a rational
transduction, then $Th(L\cap \ParikhInv{S})=U(L\cap\ParikhInv{S})$, where
$U\subseteq Z^*\times X^*$ is the rational transduction $U=\{ (v,u)\in
Z^*\times X^* \mid (v,h(u))\in T \}$.  We may assume that $X\cap Z=\emptyset$.
Construct a regular language $R\subseteq (X\cup Z)^*$ with
$U=\{(\pi_Z(w),\pi_X(w)) \mid w\in R\}$. With this, we have
\begin{align*}
U(L\cap \ParikhInv{S}) &= \pi_Z\left((R\cap (L\shuffle Z^*))\cap\ParikhInv{S+Z^\oplus}\right).
\end{align*}
Since $\C$ is an effective full semi-AFL, and thus $R\cap (L\shuffle Z^*)$ is
effectively in $\C$, the right hand side is effectively contained in
$\HomSLI{\C}$. This proves that $\HomSLI{\C}$ is an effective full trio.

Let us prove effective closure under union. Now suppose $L_i\subseteq X_i^*$,
$S_i\subseteq X_i^\oplus$, and $h_i\colon X_i^*\to Y^*$ for $i=1,2$.  If
$\bar{X}_2$ is a disjoint copy of $X_2$ with bijection $\varphi\colon X_2\to
\bar{X}_2$, then
\[ h_1(L_1\cap\ParikhInv{S_1})\cup h_2(L_2\cap\ParikhInv{S_2})=h((L_1\cup \varphi(L_2))\cap \ParikhInv{S_1\cup \varphi(S_2)}), \]
where $h\colon X_1\cup \bar{X}_2\to Y$ is the map with $h(x)=h_1(x)$ for $x\in
X_1$ and $h(x)=h_2(\varphi(x))$ for $x\in \bar{X}_2$. This proves that
$\HomSLI{\C}$ is effectively closed under union.

It remains to be shown that $\HomSLI{\C}$ is Presburger closed. Suppose
$L\in\C$, $L\subseteq X^*$, $S\subseteq X^\oplus$ is semilinear, $h\colon
X^*\to Y^*$ is a morphism, and $T\subseteq Y^\oplus$ is another semilinear set.
Let $\varphi\colon X^\oplus\to Y^\oplus$ be the morphism with
$\varphi(\Parikh{w})=\Parikh{h(w)}$ for every $w\in X^*$. Moreover, consider
the set
\[ T'=\{\mu\in X^\oplus \mid \varphi(w)\in T \}=\{\Parikh{w}\mid w\in X^*,~\Parikh{h(w)}\in T\}. \]
It is clearly Presburger definable in terms of $T$ and hence effectively semilinear.
Furthermore, we have
\begin{align*}
h(L\cap\ParikhInv{S})\cap\ParikhInv{T}=h(L\cap \ParikhInv{S\cap T'}).
\end{align*}
This proves that $\HomSLI{\C}$ is effectively Presburger closed.
\end{proof}

\begin{proof}[Proof of \cref{hierarchy:closure}]
\Cref{hierarchy:closure} follows from
\cref{basics:alg:semiafl,basics:sli:presburger:trio}.  The uniform algorithm
recursively applies the transformations described therein.
\end{proof}

\section{Proof of \cref{basics:fsemilinear}}

\begin{cproposition}\label{basics:sli:semilinear}
If $\C$ is  semilinear, then so is $\HomSLI{\C}$.
Moreover, if $\C$ is effectively semilinear, then so is $\HomSLI{\C}$.
\end{cproposition}
\begin{proof}
Since morphisms effectively preserve semilinearity, it suffices to show that
$\Parikh{L\cap\ParikhInv{S}}$ is (effectively) semilinear for each $L\in\C$,
$L\subseteq X^*$, and semilinear $S\subseteq X^\oplus$.  This, however, is easy
to see since $\Parikh{L\cap\ParikhInv{S}}=\Parikh{L}\cap S$ and the semilinear
subsets of $X^\oplus$ are closed under intersection (they coincide with the
Presburger definable sets). Furthermore, if a semilinear representation of
$\Parikh{L}$ can be computed, this is also the case for $\Parikh{L}\cap S$.
\end{proof}

\begin{proof}[Proof of \cref{basics:fsemilinear}]
The semilinearity follows from \cref{basics:sli:semilinear} and a result by
van~Leeuwen~\cite{vanLeeuwen1974}, stating that if $\C$ is semilinear, then so
is $\Alg{\C}$.

The computation of (semilinear representations of) Parikh images can be done
recursively.  The procedure in \cref{basics:sli:semilinear} describes the
computation for languages in $\HF_i$.  In order to compute the Parikh image of
a language in $\HG_i=\Alg{\HF_i}$, consider an $\HF_i$-grammar $G$. Replacing
each right-hand side by a Parikh equivalent regular language yields a
$\Reg$-grammar $G'$ that is Parikh equivalent to $G$. Since $G'$ is effectively
context-free, one can compute the Parikh image for $G'$.
\end{proof}

\section{Simple constructions of PAIM}
This section contains simple lemmas for the construction of PAIM.

\begin{clemma}[Unions]\label{pa:union}
Given $i\in\N$ and languages $L_0,L_1\in\HG_i$, along with a PAIM in $\HG_i$
for each of them, one can construct a PAIM for $L_0\cup L_1$ in $\HG_i$.
\end{clemma}
\begin{proof}
One can find a PAIM $(K^{(i)}, C^{(i)}, P^{(i)}, (P^{(i)}_c)_{c\in C^{(i)}},
\varphi^{(i)}, \marker)$ for $L_i$ in $\C$ for $i=0,1$ such that $C^{(0)}\cap
C^{(1)}=P^{(0)}\cap P^{(1)}=\emptyset$. Then $K=K^{(0)}\cup K^{(1)}$ is
effectively contained in $\HG_i$ and can be turned into a PAIM $(K,C,P,(P_c)_{c\in
c},\varphi,\marker)$ for $L_0\cup L_1$.
\end{proof}

\begin{clemma}[Homomorphic images]\label{pa:morphism}
Let $h\colon X^*\to Y^*$ be a morphism. Given $i\in\N$ and a PAIM for
$L\in\HG_i$ in $\HG_i$, one can construct a PAIM for $h(L)$ in $\HG_i$.
\end{clemma}
\begin{proof}
Let $(K,C,P,(P_c)_{c\in C},\varphi,\marker)$ be a PAIM for $L$ and let
$\bar{h}\colon X^\oplus\to Y^\oplus$ be the morphism with
$\bar{h}(x)=\Parikh{h(x)}$ for $x\in X$.  Define the new morphism
$\varphi'\colon (C\cup P)^\oplus\to Y^\oplus$ by
$\varphi'(\mu)=\bar{h}(\varphi(\mu))$. Moreover, let $g\colon (C\cup X\cup
P\cup \{\marker\})^*\to (C\cup Y\cup P\cup \{\marker\})^*$ be the extension of
$h$ that fixes $C\cup P\cup \{\marker\}$.  Then $(g(K),C,P,(P_c)_{c\in
C},\varphi',\marker)$ is clearly a PAIM for $h(L)$ in $\HG_i$.
\end{proof}

\begin{clemma}[Linear decomposition]\label{pa:decomposelinear}
Given $i\in\N$ and $L\in\HG_i$ along with a PAIM in $\HG_i$, one can construct
$L_1,\ldots,L_n\in\HG_i$, each together with a linear PAIM in $\HG_i$, such
that $L=L_1\cup\cdots\cup L_n$.
\end{clemma}
\begin{proof}
Let $(K,C,P,(P_c)_{c\in C},\varphi,\marker)$ be a PAIM for $L\subseteq X^*$.
For each $c\in C$, let $K_c=K\cap c(X\cup P\cup\{\marker\})^*$.  Then $(K_c,
\{c\}, P_c, P_c, \varphi_c,\marker)$, where $\varphi_c$ is the restriction of
$\varphi$ to $(\{c\}\cup P_c)^\oplus$, is a PAIM for $\pi_X(K_c)$ in
$\HG_i$.  Furthermore, $L=\bigcup_{c\in C}\pi_X(K_c)$.
\end{proof}

\begin{clemma}[Presence check]\label{pa:checkif}
Let $X$ be an alphabet and $x\in X$. Given $i\in\N$ and a PAIM for $L\subseteq
X^*$ in $\HG_i$, one can construct a PAIM for $L\cap X^*xX^*$ in $\HG_i$.
\end{clemma}
\begin{proof}
Since
\[ (L_1\cup\cdots\cup L_n)\cap X^*xX^* = (L_1\cap X^*xX^*)\cup\cdots\cup (L_n\cap X^*xX^*),\]
\cref{pa:decomposelinear} and \cref{pa:union} imply that we may assume that the
PAIM $(K,C,P,(P_c)_{c\in C},\varphi,\marker)$ for $L$ is linear, say $C=\{c\}$
and $P=P_c$.  Since in the case $\varphi(c)(x)\ge 1$, we have $L\cap X^*xX^*=L$
and there is nothing to do, we assume $\varphi(c)(x)=0$.

Let $C'=\{(c,p)\mid p\in P, \varphi(p)(x)\ge 1\}$ be a new alphabet and let
\[ K'=\{(c,p)uv \mid (c,p)\in C',~u,v\in (X\cup P\cup \{\marker\})^*,~cupv\in K\}. \]
Note that $K'$ can clearly be obtained from $K$ by way of a rational
transduction and is therefore contained in $\HG_i$.
Furthermore, we let
$P'=P'_{(c,p)}=P$ and $\varphi'((c,p))=\varphi(c)+\varphi(p)$ for $(c,p)\in
C'$ and $\varphi'(p)=\varphi(p)$ for $p\in P$. Then we have
\begin{align*}
\pi_X(K')&=\{\pi_X(w) \mid w\in K,~\exists p\in P: \varphi(\pi_{C\cup P}(w))(p)\ge 1,\varphi(p)(x)\ge 1\} \\
&=\{\pi_X(w) \mid w\in K,~|\pi_X(w)|_x\ge 1\} = L\cap X^*xX^*.
\end{align*}
This proves the projection property. For each $(c,p)uv\in K'$ with $cupv\in K$, we have
\[ \varphi'(\pi_{C'\cup P'}((c,p)uv))=\varphi(\pi_{C\cup P}(cupv))=\Parikh{\pi_X(cupv)}=\Parikh{\pi_X((c,p)uv)}. \]
and thus $\varphi'(\pi_{C'\cup P'}(w))=\Parikh{\pi_X(w)}$ for every $w\in K'$.
Hence, we have established the counting property. Moreover,
\begin{align*}
\Parikh{\pi_{C'\cup P'}(K')} &= \bigcup_{p\in P} (c,p)+P'^\oplus,
\end{align*}
meaning the commutative projection property is satisfied as well.
This proves that the tuple $(\pi_{C\cup X\cup P}(K'),C',P',(P'_d)_{d\in C'},\varphi')$ is a
Parikh annotation for $L\cap X^*xX^*$ in $\HG_i$. Since
$(K,C,P,(P_c)_{c\in C},\varphi,\marker)$ is a PAIM for $L$, it follows that
$(K',C',P',(P'_d)_{d\in C'},\varphi',\marker)$ is a PAIM for $L\cap X^*xX^*$.
\end{proof}

\begin{clemma}[Absence check]\label{pa:checkifnot}
Let $X$ be an alphabet and $x\in X$. Given $i\in\N$ and a PAIM for $L\subseteq X^*$ in
$\HG_i$, one can construct a PAIM for $L\setminus X^*xX^*$ in
$\HG_i$.
\end{clemma}
\begin{proof}
Since
\[ (L_1\cup\cdots\cup L_n)\setminus X^*xX^* = (L_1\setminus X^*xX^*)\cup\cdots\cup (L_n\setminus X^*xX^*),\]
\cref{pa:decomposelinear} and \cref{pa:union} imply that we may assume that the
PAIM $(K,C,P,(P_c)_{c\in C},\varphi,\marker)$ for $L$ is linear, say $C=\{c\}$
and $P=P_c$.  Since in the case $\varphi(c)(x)\ge 1$, we have $L\setminus
X^*xX^*=\emptyset$ and there is nothing to do, we assume $\varphi(c)(x)=0$.

Let $C'=C$, $P'=P'_c=\{p\in P \mid \varphi(p)(x)=0\}$, and let
\[ K' = \{w\in K \mid |w|_p = 0~\text{for each $p\in P\setminus P'$} \}. \]
Furthermore, we let $\varphi'$ be the restriction of $\varphi$ to $(C'\cup
P')^\oplus$.  Then clearly $(K',C',(P'_c)_{c\in C'},\varphi',\marker)$ is a PAIM for
$L\setminus X^*xX^*$ in $\HG_i$.
\end{proof}

\section{Proof of \cref{nonterminal:extension}}
\begin{proof}
First, observe that there is at most one $G$-compatible extension: For each
$A\in N$, there is a $u\in T^*$ with $A\grammarsteps[G] u$ and hence
$\hat{\psi}(A)=\psi(u)$.

In order to prove existence, we claim that for each $A\in N$ and
$A\grammarsteps[G] u$ and $A\grammarsteps[G] v$ for $u,v\in T^*$, we have
$\psi(u)=\psi(v)$.  Indeed, since $G$ is reduced, there are $x,y\in T^*$ with
$S\grammarsteps[G] xAy$.  Then $xuy$ and $xvy$ are both in $\Lang{G}$ and hence
$\psi(xuy)=\psi(xvy)=h$.  In the group $H$, this implies
\[ \psi(u)=\psi(x)^{-1}h\psi(y)^{-1}=\psi(v). \]
This means a $G$-compatible extension exists: Setting $\hat{\psi}(A)=\psi(w)$
for some $w\in T^*$ with $A\grammarsteps[G] w$ does not depend on the chosen
$w$. This definition implies that whenever $u\grammarsteps[G] v$ for $u\in
(N\cup T)^*$, $v\in T^*$, we have $\hat{\psi}(u)=\hat{\psi}(v)$. Therefore, if
$u\grammarsteps[G] v$ for $u,v\in (N\cup T)^*$, picking a $w\in T^*$ with
$v\grammarsteps[G] w$ yields $\hat{\psi}(u)=\hat{\psi}(w)=\hat{\psi}(v)$.
Hence, $\hat{\psi}$ is $G$-compatible.

Now suppose $H=\Z$ and $\C=\HF_i$. Since $\Z$ is commutative, $\psi$ is
well-defined on $T^\oplus$, meaning there is a morphism $\bar{\psi}\colon
T^\oplus\to\Z$ with $\bar{\psi}(\Parikh{w})=\psi(w)$ for $w\in T^*$.  We can
therefore determine $\hat{\psi}(A)$ by computing a semilinear representation of
the Parikh image of $K=\{w\in T^* \mid A\grammarsteps[G] w\}\in\Alg{\HF_i}$
(see \cref{basics:fsemilinear}), picking an element $\mu\in\Parikh{K}$, and
compute $\hat{\psi}(A)=\bar{\psi}(\mu)$.
\end{proof}

\section{Proof of \cref{pa:matchings}}

\begin{proof}
Let $G=(N,X,P,S)$ and let $\delta\colon X^*\to\Z$ be the morphism with
$\delta(w)=\gamma_0(\pi_{X_0}(w))-\gamma_1(\pi_{X_1}(w))$ for $w\in X^*$.
Since then $\delta(w)=0$ for every $w\in \Lang{G}$, by
\cref{nonterminal:extension}, $\delta$ extends uniquely to a $G$-compatible
$\hat{\delta}\colon (N\cup X)^*\to\Z$. We claim that with
$k=\max\{|\hat{\delta}(A)| \mid A\in N\}$, each derivation tree of $G$ admits a
$k$-matching.

Consider an
$(N\cup X)$-tree $t$ and let $L_i$ be the set of $X_i$-labeled leaves.
Let $A$ be an arrow collection for $t$ and let
$d_A(\ell)$ be the number of arrows incident to $\ell\in L_0\cup L_1$.  Moreover,
let $\lambda(\ell)$ be the label of the leaf $\ell$ and let 
\[ \beta(t)=\sum_{\ell\in L_0}\gamma_0(\lambda(\ell))-\sum_{\ell\in L_1}\gamma_1(\lambda(\ell)). \]
$A$ is a \emph{partial $k$-matching} if the following holds:
\begin{enumerate}
\item if $\beta(t)\ge 0$, then $d_A(\ell)\le\gamma_0(\lambda(\ell))$ for each $\ell\in L_0$ and $d_A(\ell)=\gamma_1(\lambda(\ell)))$ for each $\ell\in L_1$.
\item if $\beta(t)\le 0$, then $d_A(\ell)\le\gamma_1(\lambda(\ell))$ for each $\ell\in L_1$ and $d_A(\ell)=\gamma_0(\lambda(\ell)))$ for each $\ell\in L_0$.
\item $d_A(s)\le k$ for every subtree $s$ of $t$.
\end{enumerate}
Hence, while in a $k$-matching the number $\gamma_i(\lambda(\ell))$ is the
degree of $\ell$ (with respect to the matching), it is merely a capacity in a partial $k$-matching.  The
first two conditions express that either all leaves in $L_0$ or all in $L_1$
(or both) are filled up to capacity, depending on which of the two sets of
leaves has less (total) capacity.

If $t$ is a derivation tree of $G$, then $\beta(t)=0$ and hence a partial
$k$-matching is already a $k$-matching.  Therefore, we show by induction on $n$
that every derivation subtree of height $n$ admits a partial $k$-matching. This
is trivial for $n=0$ and for $n>0$, consider a derivation subtree $t$ with
direct subtrees $s_1,\ldots,s_r$. Let $B$ be the label of $t$'s root and
$B_j\in N\cup X$ be the label of $s_j$'s root. Then $\hat{\delta}(B)=\beta(t)$,
$\hat{\delta}(B_j)=\beta(s_j)$ and $\beta(t)=\sum_{j=1}^r \beta(s_j)$.  By
induction, each $s_j$ admits a partial $k$-matching $A_j$. Let $A$ be the union
of the $A_j$.  Observe that since $\sum_{\ell\in L_0} d_A(\ell)=\sum_{\ell\in
L_1} d_A(\ell)$ in every arrow collection (each side equals the number of
arrows), we have
\begin{equation}\beta(t)=\underbrace{\sum_{\ell\in L_0}(\gamma_0(\lambda(\ell))-d_A(\ell))}_{=:p\ge 0}-\underbrace{\sum_{\ell\in L_1}(\gamma_1(\lambda(\ell))-d_A(\ell))}_{=:q\ge 0}. \label{eq:beta:capacity}\end{equation}
If $\beta(t)\ge 0$ and hence $p\ge q$, this equation allows us to obtain $A'$
from $A$ by adding $q$ arrows, such that each $\ell\in L_1$ has
$\gamma_1(\lambda(\ell))-d_A(\ell)$ new incident arrows. They are connected to
$X_0$-leaves so as to maintain $\gamma_0(\ell)-d_{A'}(\ell)\ge 0$.
Symmetrically, if $\beta(t)\le 0$ and hence $p\le q$, we add $p$ arrows such
that each $\ell\in L_0$ has $\gamma_0(\lambda(\ell))-d_A(\ell)$ new incident
arrows. They also are connected to $X_1$-leaves so as to maintain
$\gamma_1(\lambda(\ell))-d_{A'}(\ell)\ge 0$. Then by construction, $A'$
satisfies the first two conditions of a partial $k$-matching. Hence, it remains
to be shown that the third is fulfilled as well.

Since for each $j$, we have either $d_A(\ell)=\gamma_0(\lambda(\ell))$ for all
$\ell\in L_0\cap s_j$ or we have $d_A(\ell)=\gamma_1(\lambda(\ell))$ for all
$\ell\in L_1\cap s_j$, none of the new arrows can connect two leaves inside of
$s_j$. This means the $s_j$ are the only subtrees for which we have to verify
the third condition, which amounts to checking that $d_{A'}(s_j)\le k$ for $1\le j\le
r$.  As in \cref{eq:beta:capacity}, we have
\[ \beta(s_j)=\underbrace{\sum_{\ell\in L_0\cap s_j} (\gamma_0(\lambda(\ell))-d_A(\ell))}_{=:u\ge 0}-\underbrace{\sum_{\ell\in L_1\cap s_j} (\gamma_1(\lambda(\ell))-d_A(\ell))}_{=:v\ge 0}. \]
Since the arrows added in $A'$ have respected the capacity of each leaf, we
have $d_{A'}(s_j)\le u$ and $d_{A'}(s_j)\le v$.  Moreover, since $A_j$ is a
partial $k$-matching, we have $u=0$ or $v=0$.  In any case, we have 
$d_{A'}(s_j)\le |u-v|=|\beta(s_j)|=|\hat{\delta}(B_j)|\le k$, proving the third condition.
\end{proof}

\section{Proof of \cref{pa:lettersubstitution}}

\begin{clemma}\label{computereduced}
Given an $\HF_i$-grammar, one can compute an equivalent reduced
$\HF_i$-grammar.
\end{clemma}
\begin{proof}
Since $\HF_i$ is a Presburger closed semi-trio and has a decidable emptiness
problem, we can proceed as follows.  First, we compute the set of productive
nonterminals. We initialize $N_0=\emptyset$ and then successively compute
\[ N_{i+1}=\{A\in N \mid L\cap (N_i\cup T)^*\ne\emptyset~\text{for some $A\to L$ in $P$}\}. \]
Then at some point, $N_{i+1}=N_i$ and $N_i$ contains precisely the productive nonterminals.
Using a similar method, one can compute the set of productive nonterminals.
Hence, one can compute the set $N'\subseteq N$ of nonterminals that are
reachable and productive. The new grammar is then obtained by replacing each
production $A\to L$ with $A\to (L\cap (N'\cup T)^*)$ and removing all
productions $A\to L$ where $A\notin N'$.
\end{proof}

\begin{proof}[Proof of \cref{pa:lettersubstitution}]
In light of \cref{pa:morphism}, it clearly suffices to prove the statement in
the case that there are $a\in Z$ and $b\in Z'$ with $Z'=Z\cup \{b\}$, $b\notin
Z$ and $\sigma(x)=\{x\}$ for $x\in Z\setminus\{a\}$ and $\sigma(a)=\{a,b\}$.
Let $(K,C,P,(P_c)_{c\in C},\varphi,\marker)$ be a PAIM for $L$ in $\HG_i$.
According to \cref{computereduced}, we can assume $K$ to be given by a reduced
$\HF_i$-grammar.

We want to use \cref{consistent:substitution} to construct a PAIM for
$\sigma(L)$.  Let $X_0=Z\cup\{\marker\}$, $X_1=C\cup P$, and $\gamma_i\colon
X_i^*\to\N$ for $i=0,1$ be the morphisms with
\[ \gamma_0(w)=|w|_a,~~~~~\gamma_1(w)=\varphi(w)(a). \]
Then, by the counting property of PAIM, we have $\gamma_0(w)=\gamma_1(w)$ for each
$w\in K$. Let $Y,h$ and $Y_i,h_i,\eta_i$ be defined as in
\cref{consistent:substitution:alphabet} and
\cref{consistent:substitution:morphisms}. \Cref{consistent:substitution} allows us to construct
$\hat{K}\in\HG_i$, $\hat{K}\subseteq Y^*$, with $\hat{K}\subseteq h^{-1}(K)$,
$\pi_{X_i}(\hat{K})=\pi_{X_i}(h^{-1}(K))$ for $i=0,1$, and
$\eta_0(\pi_{X_0}(w))=\eta_1(\pi_{X_1}(w))$ for each $w\in \hat{K}$.

For each $f\in C\cup P$, let $D_f=\{(f',m)\in Y_1 \mid f'=f\}$.  With this, let
$C'=\bigcup_{c\in C} D_c$, $P'=\bigcup_{p\in P} D_p$, and
$P'_{(c,m)}=\bigcup_{p\in P_c} D_p$ for $(c,m)\in C'$.  The new morphism $\varphi'\colon
(C'\cup P')^\oplus\to Z'^\oplus$ is defined by
\begin{align*}
\varphi'((f,m))(z)&=\varphi(f)(z) & \text{for $z\in Z\setminus\{a\}$}, \\
\varphi'((f,m))(b)&=m, \\
\varphi'((f,m))(a)&=\varphi(f)(a)-m.
\end{align*}
Let $g\colon Y^*\to (C'\cup Z'\cup P'\cup \{\marker\})^*$ be the morphism
with $g((z,0))=z$ for $z\in Z$, $g((a,1))=b$, $g(x)=x$ for $x\in C'\cup
P'\cup\{\marker\}$. We claim that with $K'=g(\hat{K})$, the tuple
$(K',C',P',(P'_c)_{c\in C'},\varphi',\marker)$ is a PAIM for $\sigma(L)$.
First, note that $K'\in\HG_i$ and
\[ K'=g(\hat{K})\subseteq g(h^{-1}(K))\subseteq g(h^{-1}(C(Z\cup P)^*))\subseteq C'(Z'\cup P')^*. \]
Note that $g$ is bijective. This allows us to define $f\colon (C'\cup Z'\cup
P'\cup \{\marker\})^*\to(C\cup Z\cup P\cup\{\marker\})^*$ as the morphism
with $f(w)=h(g^{-1}(w))$ for all $w$.  Observe that then $f(a)=f(b)=a$ and
$f(z)=z$ for $z\in Z\setminus\{a,b\}$ and by the definition of $K'$, we have
$f(K')\subseteq K$ and $\sigma(L)=f^{-1}(L)$.

\begin{itemize}
\item \emph{Projection property.}
Note that $\pi_{Y_0}(u)=\pi_{Y_0}(v)$ implies $\pi_{Z'}(g(u))=\pi_{Z'}(g(v))$
for $u,v\in Y^*$. Thus, from $\pi_{Y_0}(\hat{K})=\pi_{Y_0}(h^{-1}(K))$, we
deduce
\begin{align*}
\pi_{Z'}(K')&=\pi_{Z'}(g(\hat{K}))=\pi_{Z'}(g(h^{-1}(K))) \\
&=\pi_{Z'}(f^{-1}(K))=f^{-1}(L)=\sigma(L).
\end{align*}
\item\emph{Counting property.}
Note that by the definition of $\varphi'$ and $g$, we have
\begin{align}
\varphi'(\pi_{C'\cup P'}(x))(b)=\eta_1(x)=\eta_1(g^{-1}(x)) \label{pa:lettersubst:counting:eta}
\end{align}
for every $x\in C'\cup P'$.

For $w\in K'$, we have $f(w)\in K$ and hence $\varphi(\pi_{C\cup
P}(f(w)))=\Parikh{\pi_{Z}(f(w))}$.  Since for $z\in Z\setminus\{a\}$, we have
$\varphi'(x)(z)=\varphi(f(x))(z)$ for every $x\in C'\cup P'$, it follows that
\begin{align}
\varphi'(\pi_{C'\cup P'}(w))(z)&=\varphi(\pi_{C\cup P}(f(w)))(z) \nonumber\\
&=\Parikh{\pi_Z(f(w))}(z)=\Parikh{\pi_{Z'}(w)}(z).\label{pa:lettersubst:counting:z}
\end{align}
Moreover, by \labelcref{pa:lettersubst:counting:eta} and since $g^{-1}(w)\in\hat{K}$, we have
\begin{align}
\varphi'(\pi_{C'\cup P'}(w))(b)   & =\eta_1(g^{-1}(w))=\eta_0(g^{-1}(w))=|w|_b \nonumber\\
                                  & =\Parikh{\pi_{Z'}(w)}(b).\label{pa:lettersubst:counting:b}
\end{align}
and $f(w)\in K$ yields
\begin{align*}
\varphi'(\pi_{C'\cup P'}(w))(a)+\varphi'(\pi_{C'\cup P'}(w))(b)& =\varphi(\pi_{C\cup P}(f(w)))(a) \\
                                                               & =\Parikh{\pi_{Z}(f(w))}(a) \\
                                                               & =\Parikh{\pi_{Z'}(w)}(a)+\Parikh{\pi_{Z'}(w)}(b).
\end{align*}
Together with \labelcref{pa:lettersubst:counting:b}, this implies
$\varphi'(\pi_{C'\cup P'}(w))(a)=\Parikh{\pi_{Z'}(w)}(a)$. Combining this with
\labelcref{pa:lettersubst:counting:z,pa:lettersubst:counting:b}, we
obtain $\varphi'(\pi_{C'\cup P'}(w))=\Parikh{\pi_{Z'}(w)}$.  This proves the
counting property.

\item\emph{Commutative projection property.} Observe that
\begin{align*}
\Parikh{\pi_{C'\cup P'}(K')} &= \Parikh{\pi_{Y_1}(\hat{K})}=\Parikh{\pi_{Y_1}(h^{-1}(K))} \\
                             &=\Parikh{h^{-1}(\pi_{C\cup P}(K))}=\bigcup_{c\in C'} c+P'^\oplus_c.
\end{align*}
\item\emph{Boundedness.} Since $|w|_{\marker}=|h(v)|_{\marker}$ for each $w\in
K'$ with $w=g(v)$, there is a constant bounding $|w|_{\marker}$ for $w\in K'$.

\item\emph{Insertion property.} 
Let $cw\in K'$ with $c\in C'$ and $\mu\in P'^\oplus_c$.  Then $f(\mu)\in
P_{f(c)}^\oplus$ and $f(cw)\in K$.  Write
\[ \pi_{Z'\cup\{\marker\}}(cw)=w_0\marker w_1 \marker \cdots \marker w_n \]
with $w_0,\ldots,w_n\in Z'^*$. Then
\[ \pi_{Z\cup\{\marker\}}(f(cw))=f(\pi_{Z'\cup\{\marker\}}(cw))=f(w_0)\marker\cdots\marker f(w_n). \]
By the insertion property of $K$ and since $f(cw)\in K$, there is a $v\in K$
with
\[ \pi_Z(v)=f(w_0)v_1 f(w_1)\cdots v_nf(w_n), \]
$v_1,\ldots,v_n\in Z^*$, and
$\Parikh{\pi_Z(v)}=\Parikh{\pi_Z(f(cw))}+\varphi(f(\mu))$.  In particular, we
have $\Parikh{v_1\cdots v_n}=\varphi(f(\mu))$.  Note that $\varphi'(\mu)\in
Z'^\oplus$ is obtained from $\varphi(f(\mu))\in Z^\oplus$ by replacing some
occurrences of $a$ by $b$.  Thus, by the definition of $f$, we can find words
$v'_1,\ldots,v'_n\in Z'^*$ with $f(v'_i)=v_i$ and $\Parikh{v'_1\cdots
v'_n}=\varphi'(\mu)$.  Then the word
\[ w' = w_0v'_1w_1\cdots v'_nw_n\in Z'^* \]
statisfies $\pi_{Z'\cup\{\marker\}}(cw)\preceq_{\marker} w'$,
$\Parikh{w'}=\Parikh{\pi_{Z'}(cw)}+\varphi'(\mu)$ and
\[ f(w')=f(w_0)v_1f(w_1)\cdots v_nf(w_n)=\pi_Z(v)\in \pi_Z(K)=L. \]
Since $f^{-1}(L)=\sigma(L)$, this means $w'\in\sigma(L)$.  We have thus
established the insertion property.
\end{itemize}
We conclude that the tuple $(K',C',P',(P'_c)_{c\in C'},\varphi',\marker)$ is a
PAIM in $\HG_i$ for $\sigma(L)$.
\end{proof}

\section{Proof of \cref{pa:substitution}}
\begin{proof}
Let $\sigma\colon X^*\to\Powerset{Y^*}$.
Assuming that for some $a\in X$, we have $\sigma(x)=\{x\}$ for all $x\in
X\setminus\{a\}$ means no loss of generality. According to
\cref{pa:morphism}, we may also assume that $\sigma(a)\subseteq Z^*$ for some alphabet $Z$ with
$Y=X\uplus Z$. If $\sigma(a)=L_1\cup \cdots\cup L_n$, then first substituting
$a$ by $\{a_1,\ldots,a_n\}$ and then each $a_i$ by $L_i$ has the same effect as
applying $\sigma$. Hence, \cref{pa:lettersubstitution} allows us to assume
further that the PAIM given for $\sigma(a)$ is linear. Finally, since
$\sigma(L)=(L\setminus X^*aX^*)\cup \sigma(L\cap X^*aX^*)$,
\cref{pa:checkif,pa:checkifnot,pa:union} imply that we may also assume
$L\subseteq X^*aX^*$.

Let $(K,C,P,(P_c)_{c\in C},\varphi,\marker)$ be a PAIM for $L$ and
$(\hat{K},\hat{c},\hat{P},\hat{\varphi},\marker)$ be a linear PAIM for
$\sigma(a)$.  The idea of the construction is to replace each occurrence of $a$
in $K$ by words from $\hat{K}$ after removing $\hat{c}$. However, in order to
guarantee a finite bound for the number of occurrences of $\marker$ in the
resulting words, we also remove $\marker$ from all but one inserted words from
$\hat{K}$. The new map $\varphi'$ is then set up to so that if $f\in C\cup P$
represented $m$ occurrences of $a$, then $\varphi'(f)$ will represent $m$ times
$\hat{\varphi}(\hat{c})$.

Let $C'=C$, $P'_c=P_c\cup\hat{P}$, $P'=\bigcup_{c\in C'}P'_c$, and
$\varphi'\colon (C'\cup P')^\oplus\to Y^\oplus$ be the morphism with
\begin{align*}
\varphi'(f)&=\varphi(f)~-~\varphi(f)(a)\cdot a ~+~ \varphi(f)(a)\cdot\hat{\varphi}(\hat{c}) & & \text{for $f\in C\cup P$}, \\
\varphi'(f)&=\hat{\varphi}(f) & & \text{for $f\in \hat{P}$}.
\end{align*}
Let $a_{\marker}$ be a new symbol and 
\[ \bar{K}=\{ua_{\marker}v \mid uav\in K,~|u|_a=0 \}. \]
In other words, $\bar{K}$ is obtained by replacing in each word from $K$
the first occurrence of $a$ with $a_{\marker}$. The occurrence of
$a_\marker$ will be the one that is replaced by all of $\hat{K}$, the
occurrences of $a$ are replaced by $\pi_{\{\hat{c}\}\cup
Z\cup\hat{P}}(\hat{K})$. Let $\tau$ be the substitution
\begin{align*}
\tau\colon (C\cup X\cup P\cup \{\marker,a_{\marker}\})^*        &\longrightarrow\Powerset{(C'\cup Z\cup P'\cup\{\marker\})^*}   & \\
x                                                               &\longmapsto\{x\},~~\text{for $x\in C\cup X\cup P\cup\{\marker\}$, $x\ne a$}, \\
a_{\marker}                                                     &\longmapsto\pi_{Z\cup \hat{P}\cup\{\marker\}}(\hat{K}),  \\
a&\longmapsto\pi_{Z\cup \hat{P}}(\hat{K}).              &
\end{align*}
We claim that with $K'=\tau(\bar{K})$, the tuple $(K',C',P',(P'_c)_{c\in
C'},\varphi',\marker)$ is a PAIM in $\HG_i$ for $\sigma(L)$.  First,
since $\HG_i$ is closed under rational transductions and substitutions,
$K'$ is in $\HG_i$.

\begin{itemize}
\item\emph{Projection property.}
Since $L=\pi_X(K)$ and $\sigma(a)=\pi_Z(\hat{K})$,
we have $\sigma(L)=\pi_Z(K')$.
\item\emph{Counting property.}
Let $w\in K'$. Then there is a $u=cu_0au_1\cdots au_n\in K$, $u_i\in (C\cup
X\cup P\cup \{\marker\})^*$, $c\in C$, and $|u_i|_a=0$ for $i=0,\ldots,n$ and
$w=cu_0w_1u_1\cdots w_nu_n$ with $w_1\in\pi_{Z\cup
\hat{P}\cup\{\marker\}}(\hat{K})$, $w_i\in \pi_{Z\cup \hat{P}}(\hat{K})$ for
$i=2,\ldots,n$. This means
\begin{equation}\Parikh{\pi_Z(w_i)}=\hat{\varphi}(\hat{c})+\hat{\varphi}(\pi_{\hat{P}}(w_i)). \label{pa:substitution:a}\end{equation}
Since $\varphi(\pi_{C\cup P}(u))(a)=\Parikh{\pi_X(u)}=n$, we have
\begin{align}
\varphi'(\pi_{C'\cup P'}(u)) &= \varphi(\pi_{C\cup P}(u)) - n\cdot a + n\cdot\hat{\varphi}(\hat{c}) \label{pa:substitution:b}\\
&=\Parikh{\pi_X(u)}-n\cdot a+n\cdot\hat{\varphi}(\hat{c}). \nonumber
\end{align}
\Cref{pa:substitution:a,pa:substitution:b} together imply
\begin{align*}
\varphi'(\pi_{C'\cup P'}(w)) &=\varphi'(\pi_{C'\cup P'}(u))+\sum_{i=1}^n\varphi'(\pi_{P'}(w_i)) \\
        &=\Parikh{\pi_X(u)}-n\cdot a + n\cdot \hat{\varphi}(\hat{c})+\sum_{i=1}^n \left(\Parikh{\pi_Z(w_i)}-\hat{\varphi}(\hat{c})\right) \\
        &=\Parikh{\pi_X(u)}-n\cdot a + \sum_{i=1}^n \Parikh{\pi_Z(w_i)}=\Parikh{\pi_Z(w)}.
\end{align*}
\item\emph{Commutative projection property.}
Let $c\in C'$ and $\mu\in P'^\oplus_c$ and write $\mu=\nu+\hat{\nu}$ with
$\nu\in P_c^\oplus$ and $\hat{\nu}\in \hat{P}^\oplus$. Then there is a $cw\in
K$ with $\Parikh{\pi_{C\cup P}(cw)}=c+\nu$. Since $L\subseteq X^*aX^*$, we can
write $w=cu_0au_1\cdots au_n$ with $|u_i|_a=0$  for $0\le i\le n$ and $n\ge
1$.  Moreover, there are $\hat{c}\hat{w}\in \hat{K}$ and
$\hat{c}\hat{w}'\in\hat{K}$ with
$\Parikh{\pi_{\{\hat{c}\cup\hat{P}}(\hat{c}\hat{w})}=\hat{c}+\hat{\nu}$ and
$\Parikh{\pi_{\hat{c}\cup\hat{P}}(\hat{c}\hat{w}')}=\hat{c}$.  By definition of $K'$, the word
\[ w'=cu_0\hat{w}u_1\hat{w}'u_2\cdots \hat{w}'u_n \]
is in $K'$ and satisfies $\Parikh{\pi_{C'\cup P'}(w')}=c+\nu+\hat{\nu}=c+\mu$.
This proves
\[ \bigcup_{c\in C'} c+P'^\oplus_c\subseteq\Parikh{\pi_{C'\cup P'}(K')}. \]
The other inclusion is clear by definition.  We have thus established that
the tuple $(\pi_{C'\cup Z\cup P'}(K'), C', P', (P'_c)_{c\in
C'},\varphi')$ is a Parikh annotation in $\HG_i$ for $\sigma(L)$.
\item\emph{Boundedness.}
Note that if $|w|_{\marker}\le k$ for all $w\in K$ and $|\hat{w}|_{\marker}\le
\ell$ for all $\hat{w}\in \hat{K}$, then $|w'|_{\marker}\le k+\ell$ for all
$w'\in K'$ by construction of $K'$, implying boundedness.
\item\emph{Insertion property.} The insertion property follows from the
insertion property of $K$ and $\hat{K}$.
\end{itemize}
\end{proof}

\section{Proof of \cref{pa:onenonterminal}}
\begin{clemma}[Sentential forms]\label{pa:onenonterminal:sf}
Let $G=(N,T,P,S)$ be an $\HG_i$-grammar with $N=\{S\}$, $P=\{S\to L\}$,
and $L\subseteq (N\cup T)^*S(N\cup T)^*$.  Furthermore, suppose a PAIM in
$\HG_i$ is given for $L$. Then one can construct a PAIM in $\HG_i$
for $\SententialForms{G}$.
\end{clemma}
\begin{proof}
Observe that applying the production $S\to L$ with $w\in L$ contributes
$\Parikh{w}-S$ to the Parikh image of the sentential form.  Therefore, we have
$\Parikh{\SententialForms{G}}=S+(\Parikh{L}-S)^\oplus$ and we can construct a
PAIM for $\SententialForms{G}$ using an idea to obtain a semilinear
representation of $U^\oplus$ for semilinear sets $U$.  If $U=\bigcup_{j=1}^n
\mu_j+F_j^\oplus$ for $\mu_j\in X^\oplus$ and finite $F_j\subseteq X^\oplus$,
then
\[ U^\oplus = \bigcup_{D\subseteq\{1,\ldots,n\}} \sum_{j\in D} \mu_j + \left(\bigcup_{j\in D} \{\mu_j\}\cup F_j\right)^\oplus.\]
The symbols representing constant and period vectors for $\SententialForms{G}$
are therefore set up as follows.  Let $(K,C,P,(P_c)_{c\in C},\varphi,\marker)$
be a PAIM for $L$ in $\HG_i$.  and let $S'$ and $S_D$ and $d_D$ be new symbols
for each $D\subseteq C$.  Moreover, let $C'=\{d_D \mid D\subseteq C\}$ and
$P'=C\cup P$ with $P'_{d_D}=D\cup\bigcup_{c\in D}P_c$. 
We will use the shorthand $X=N\cup T$.  Observe that since $L\subseteq
X^*SX^*$, we have $\varphi(c)(S)\ge 1$ for each $c\in C$.
We can therefore define the
morphism $\varphi'\colon (C'\cup P')^\oplus\to X^\oplus$ as
\begin{align}
\varphi'(p) &= \varphi(p)    & &\text{for $p\in P$,} \nonumber\\
\varphi'(c) &= \varphi(c)-S  & &\text{for $c\in C$,}\label{pa:onenon:constantrep} \\
\varphi'(d_D) &= S+\sum_{c\in D}\varphi'(c). & &\label{pa:onenon:newconstantrep}
\end{align}
The essential idea in our construction is to use modified versions of $K$
as right-hand-sides of a grammar. These modified versions are obtained as follows.
For each $D\subseteq C$, we define the rational transduction $\delta_D$ which
maps each word $w_0Sw_1\cdots Sw_n\in (C\cup X\cup P\cup\{\marker\})^*$,
$|w_i|_S=0$ for $0\le i\le n$, to all words $w_0 S_{D_1}w_1\cdots S_{D_n}w_n$
for which
\begin{align*} &D_1\cup\cdots\cup D_n=D,  &    &D_i\cap D_j=\emptyset~\text{for $i\ne j$.} \end{align*}
Thus, $\delta_D$ can be thought of as distributing the elements of $D$ among
the occurrences of $S$ in the input word.  The modified versions of $K$ are
then given by
\begin{align*}
&K_D = \delta_D(\pi_{C\cup X\cup P}(K)),   &   &K_D^{c} = \delta_{D\setminus c}(c^{-1}K).
\end{align*}

In the new annotation, the symbol $d_D$ represents $S+\sum_{c\in
D}(\varphi(c)-S)$.  Since each symbol $c\in C$ still represents $\varphi(c)-S$,
we cannot insert a whole word from $K$ for each inserted word from $L$: This would
insert a $c\in C$ in each step and we would count $\sum_{c\in D}(\varphi(c)-S)$ twice. Hence, in order to compensate for the new
constant symbol $d_D$, when generating a word starting with $d_D$,
we have to prevent exactly one occurrence of $c$ for each $c\in D$ from appearing.
To this end, we use the nonterminal $S_D$, which only allows derivation
subtrees in which of each $c\in D$, precisely one occurrence has been left out, i.e.
a production $S_D\to K_D^c$ (for some $D\subseteq C$) has been applied. In the productions $S_D\to K_D$
the symbol from $C$ on the right hand side is allowed to appear.

In order to have only a bounded number of occurrences of $\marker$, one of our
modified versions of $K$ (namely $K_D^c$) introduces $\marker$  and the other
one ($K_D$) does not. Since when generating a word starting with $d_D$, our
grammar makes sure that for each $c\in D$, a production of the form $S_E\to
K_E^c$ is used precisely once (and otherwise $S_E\to K_E$), the set $K_E^c$ is
set up to contain $\marker$.  This will guarantee that during the insertion
process simulating $S\to L$, we insert at most $|C|\cdot \ell$ occurrences of
$\marker$, where $\ell$ is an upper bound for $|w|_{\marker}$ for $w\in K$.

Let $N'=\{S'\}\cup \{S_D\mid D\subseteq C\}$ and let $\hat{P}$ consist of the
following productions:
\begin{align}
S' &\to \{d_D\marker S_D\marker\mid D\subseteq C\}  & & \label{pa:one:production:first} \\
S_\emptyset &\to \{S\} & & \label{pa:one:production:last} \\
S_D &\to K_D & & \text{for each $D\subseteq C$} \label{pa:one:production:withoutmarker} \\
S_D&\to K_D^{c} & & \text{for each $D\subseteq C$ and $c\in D$.} \label{pa:one:production:withmarker}
\end{align}
Finally, let $M$ be the regular language
\[ M=\bigcup_{D\subseteq C} \{ w\in (C'\cup X\cup P'\cup \{\marker\})^* \mid \pi_{C'\cup P'}(w)\in d_D P'^*_{d_D} \}. \]
By intersecting with $M$, we make sure that the commutative projection property is satisfied.
We shall prove that with the grammar $G'=(N',C'\cup X\cup P'\cup
\{\marker\},\hat{P},S')$ and $K'=\Lang{G'}\cap M$, the tuple
$(K',C',P',(P'_c)_{c\in C'},\varphi',\marker)$ is a PAIM for
$\SententialForms{G}$ in $\HG_i$. By definition, $\Lang{G'}$ is contained in
$\Alg{\HG_i}=\HG_i$ and hence $K'$ since $\HG_i$ is a full semi-AFL.

\newcommand{\constantsetname}{\rho}
\newcommand{\constantset}[1]{\constantsetname(#1)}
Let $h\colon (N'\cup C'\cup X\cup P'\cup\{\marker\})^*\to (C'\cup X\cup P'\cup \{\marker\})^*$ be the morphism that fixes $C'\cup X\cup P'\cup \{\marker\}$ and
satisfies $h(S')=h(S_D)=S$ for $D\subseteq C$.  Moreover, regard $\Powerset{C}$
as a monoid with $\cup$ as its operation. Then $\constantsetname\colon (N'\cup
X)^*\to\Powerset{C}$ is the morphism with $\constantset{S_D}=D$ and
$\constantset{S'}=\constantset{x}=\emptyset$ for $x\in X$.  Furthermore, let
$|w|_\marker \le \ell$ for all $w\in K$. We claim that for each $n\ge 0$,
$d_D\marker S_D\marker\grammarstepsn[G']{n} w$ implies
\begin{enumerate}
\item\label{pa:one:disjoint}    if $w=u_0S_{D_1}u_1\cdots S_{D_n}u_n$ with $u_i\in X^*$ for $0\le i\le n$, then $D_i\cap D_j=\emptyset$ for $i\ne j$.
\item\label{pa:one:projection}  $h(\pi_{N'\cup X}(w))\in \SententialForms{G}$,
\item\label{pa:one:counting}    $\varphi'(\pi_{C'\cup P'}(w))=\Parikh{h(\pi_{N'\cup X}(w))}+\sum_{c\in\constantset{w}} \varphi'(c)$,
\item\label{pa:one:boundedness} $|w|_\marker \le 2+|D\setminus\constantset{w}|\cdot \ell$, and
\item\label{pa:one:insertion}   for each $\mu\in \left(D\cup \bigcup_{c\in
D\setminus\constantset{w}} P_c\right)^\oplus$, there is a $w'\in
\SententialForms{G}$ such that $h(\pi_{N'\cup X\cup \{\marker\}}(w))
\preceq_{\marker} w'$ and $\Parikh{w'}=\Parikh{h(\pi_{N'\cup X}(w))}+\varphi'(\mu)$.
\end{enumerate}
We establish this claim using induction on $n$.  Observe that all these
conditions are satisfied in the case $n=0$, i.e.  $w=d_D\marker S_D \marker$,
\cref{pa:one:disjoint,pa:one:projection,pa:one:counting,pa:one:boundedness}
follow directly by distinguishing among the productions in $G'$. Therefore, we
only prove \cref{pa:one:insertion} in the induction step.

Suppose $n>0$ and $d_D\marker
S_D\marker\grammarstepsn[G']{n-1}\bar{w}\grammarstep[G'] w$.  If the production
applied in $\bar{w}\grammarstep[G'] w$ is $S_\emptyset\to\{S\}$, then
$\constantset{w}=\constantset{\bar{w}}$ and $h(\pi_{N'\cup X\cup
\{\marker\}})(w)=h(\pi_{N'\cup X\cup\{\marker\}})(\bar{w})$, so that
\cref{pa:one:insertion} follows immediately from the same condition for
$\bar{w}$. If the applied production is of the form
\labelcref{pa:one:production:withoutmarker} or \labelcref{pa:one:production:withmarker}, then we have
$\constantset{w}\subseteq\constantset{\bar{w}}$ and hence $\constantset{\bar{w}}=\constantset{w}\cup E$ for some $E\subseteq D$, $|E|\le 1$.
Then
\[ \bigcup_{c\in D\setminus\constantset{w}} P_c=\bigcup_{c\in D\setminus\constantset{\bar{w}}} P_c \cup \bigcup_{c\in E} P_c. \]
We can therefore decompose $\mu\in \left(D\cup \bigcup_{c\in
D\setminus\constantset{w}} P_c\right)^\oplus$ into $\mu=\bar{\mu}+\nu$ with
$\bar{\mu}\in \left(D\cup \bigcup_{c\in D\setminus\constantset{\bar{w}}}
P_c\right)^\oplus$ and $\nu\in \left(\bigcup_{c\in E} P_c\right)^\oplus$. By
induction, we find a $\bar{w}'\in\SententialForms{G}$ such that
$h(\pi_{N'\cup X\cup\{\marker\}}(\bar{w}))\preceq_{\marker}\bar{w}'$
and $\Parikh{\bar{w}'}=\Parikh{h(\pi_{N'\cup X}(\bar{w}))}+\varphi'(\bar{\mu})$.
Let $\bar{w}=xSy$ be the decomposition facilitating the step $\bar{w}\grammarstep[G'] w$ and let $w=xzy$.
\begin{itemize}
\item If the production applied in $\bar{w}\grammarstep[G'] w$ is of the form
\labelcref{pa:one:production:withoutmarker}.  Then
$\constantset{w}=\constantset{\bar{w}}$ and hence $E=\emptyset$ and $\nu=0$.
Furthermore, $z\in K_F$ for some $F\subseteq C$.
We define $z'=h(\pi_{N'\cup X}(z))$. Note that then $z'\in \pi_{X}(K)=L$ and $\Parikh{z'}=\Parikh{\pi_X(z)}+\varphi'(\nu)$.
\item If the production applied in $\bar{w}\grammarstep[G'] w$ is of the form
\labelcref{pa:one:production:withmarker}. Then $z\in K_F^c$ for some $c\in
F\subseteq C$ and thus $h(z)\in c^{-1}K$. This implies
$\constantset{\bar{w}}=\constantset{w}\cup \{c\}$, $E=\{c\}$, and hence $\nu\in
P_c^\oplus$. The insertion property of $K$ provides a $z'\in L$ such that
$\pi_{X\cup \{\marker\}}(h(z)) \preceq_{\marker} z'$ and
$\Parikh{z'}=\Parikh{\pi_{X}(h(z))}+\varphi(\nu)=\Parikh{\pi_X(h(z))}+\varphi'(\nu)$.
\end{itemize}
In any case, we have 
\begin{align*}&z'\in L,  &  &h(\pi_{N'\cup X\cup\{\marker\}}(z))\preceq_{\marker} z', & &\Parikh{z'}=\Parikh{h(\pi_{N'\cup X}(z))}+\varphi'(\nu). \end{align*}
Recall that $\bar{w}=xSy$ and $w=xzy$. Since $\bar{w}\preceq_{\marker} \bar{w}'$, we can find $x',y'$ with
\begin{align*}&\bar{w}'=x'Sy',  & & h(\pi_{N'\cup X}(x))\preceq_{\marker} x', & & h(\pi_{N'\cup X}(y))\preceq_{\marker} y'.\end{align*}
Choose $w' = x'z'y'$. Then $\SententialForms{G}\ni\bar{w}'\grammarstep[G] w'$ and thus $w'\in\SententialForms{G}$. Moreover,
\begin{align*}
h(\pi_{N'\cup X\cup\{\marker\}}(w)) &= h(\pi_{N'\cup X\cup \{\marker\}}(x))h(\pi_{N'\cup X\cup \{\marker\}}(z))h(\pi_{N'\cup X\cup \{\marker\}}(y)) \\
&\preceq_{\marker} x'z'y'=w'.
\end{align*}
Finally, $w'$ has the desired Parikh image:
\begin{align*}
\Parikh{w'}&=\Parikh{\bar{w}'}-S+\Parikh{z'} \\
&=\Parikh{h(\pi_{N'\cup X}(\bar{w}))}+\varphi'(\bar{\mu})-S+\Parikh{z'} \\
&=\Parikh{h(\pi_{N'\cup X}(\bar{w}))}+\varphi'(\bar{\mu})-S+\Parikh{h(\pi_{N'\cup X}(z))}+\varphi'(\nu) \\
&=\Parikh{h(\pi_{N'\cup X}(w))}+\varphi'(\bar{\mu})+\varphi'(\nu) \\
&=\Parikh{h(\pi_{N'\cup X}(w))}+\varphi'(\mu).
\end{align*}
This completes the induction step for \cref{pa:one:insertion}.

We now use our claim to prove that we have indeed constructed a PAIM.
\begin{itemize}
\item \emph{Projection property}. Our claim already entails $\pi_X(K')\subseteq\SententialForms{G}$:
For $w\in (C'\cup X\cup P'\cup \{\marker\})^*$ with $d_D\marker S_D\marker\grammarsteps[G'] w$,
we have $\pi_X(w)=h(\pi_{N'\cup X}(w))\in\SententialForms{G}$ by \cref{pa:one:projection}.
In order to prove $\SententialForms{G}\subseteq\pi_X(K')$, suppose
$w\in\SententialForms{G}$ and let $t$ be a partial derivation tree for $G$ with
root label $S$ and $\yield{t}=w$. 
Since $\children{x}\in L$ for each inner node $x$ of $t$, we can find a
$c_xw_x\in K$ with $\pi_X(c_xw_x)=\children{x}$.  Then in particular
$\children{x}\preceq c_xw_x$, meaning we can obtain a tree $t'$ from $t$ as
follows: For each inner node $x$ of $t$, add new leaves directly below $x$ so
as to have $c_xw_x$ as the new sequence of child labels of $x$. Note that the
set of inner nodes of $t'$ is identical to the one of $t$. Moreover, we have
$\pi_X(\yield{t'})=w$.

Let $D=\{c_x \mid \text{$x$ is an inner node in $t'$}\}$.  We pick for each
$c\in D$ exactly one inner node $x$ in $t'$ such that $c_x=c$; we denote the
resulting set of nodes by $R$. We now obtain $t''$ from $t'$ as follows: For
each $x\in R$, we remove its $c_x$-labeled child; for each $x\notin R$, we
remove all $\marker$-labeled children. Note that again, the inner nodes of $t''$
are the same as in $t$ and $t'$. Moreover, we still have $\pi_X(\yield{t''})=w$.

For each inner node $x$ in $t''$, let $D_x=\{c_y \mid \text{$y\in R$ is below
$x$ in $t''$}\}$. Note that in $t,t',t''$, every inner node has the label $S$.
We obtain the tree $t'''$ from $t''$ as follows. For each inner node $x$ in
$t''$, we replace its label $S$ by $S_{D_x}$. Then we have
$\pi_X(h(\yield{t'''}))=w$.  Clearly, the root node of $t'''$ is labeled $S_D$.
Furthermore, the definition of $K_E$ and $K_E^c$ yields that $t'''$ is a
partial derivation tree for $G'$. Hence
\[ S'~~\grammarstep[G']~~d_D\marker S_D\marker~~\grammarsteps[G']~~d_D\marker\yield{t'''}\marker. \]
Since in $t'''$, every leaf has a label in $T\cup\{S_\emptyset\}$, we have
$S'\grammarsteps[G'] d_D \marker h(\yield{t'''})\marker$. This means
$d_D\marker h(\yield{t'''})\marker\in \Lang{G'}$. Furthermore, we clearly have
$d_D\marker h(\yield{t'''})\marker\in M$ and since $\pi_X(d_D\marker
h(\yield{t'''})\marker)=w$, this implies $w\in \pi_X(K')$.
\item \emph{Counting property}. Apply \cref{pa:one:counting} in our claim to a
word $w\in (C'\cup X\cup P'\cup \{\marker\})^*$ with $d_D\marker S_D\marker\grammarsteps[G'] w$.  Since
$\constantset{w}=\emptyset$ and $h(\pi_{N'\cup X}(w))=\pi_X(w)$, this yields
$\varphi'(\pi_{C'\cup P'}(w))=\Parikh{\pi_X(w)}$.
\item \emph{Commutative projection property}. 
Since $K'\subseteq M$, we clearly have $\Parikh{\pi_{C'\cup P'}(K')}\subseteq\bigcup_{c\in C'}
c+P'^\oplus_c$.

For the other inclusion, let $D\subseteq C$ with $D=\{c_1,\ldots,c_n\}$.
Suppose $\mu\in \bigcup_{c\in C'} c+P'^\oplus_c$,
$\mu=d_D+\nu+\sum_{i=1}^n\xi_i$ with $\nu\in D^\oplus$ and $\xi_i\in
P_{c_i}^\oplus$ for $1\le i\le n$.

The commutative projection property of $K$ allows us to choose for $1\le i\le n$ words $u_i,v_i\in
K$ such that
\begin{align*} && \Parikh{\pi_{C\cup P}(u_i)}=c_i, && \Parikh{\pi_{C\cup P}(v_i)}=c_i+\xi_i.\end{align*}
The words $v'_0,\ldots,v'_n$ are constructed as follows. Let $v'_0=d_D \marker
S\marker$ and let $v'_{i+1}$ be obtained from $v'_i$ by replacing the first
occurrence of $S$ by $c_{i+1}^{-1}v_{i+1}$. Furthermore, let $v''_i$ be
obtained from $v'_i$ by replacing the first occurrence of $S$ by
$S_{\{c_{i+1},\ldots,c_n\}}$ and all other occurrences by $S_\emptyset$.  Then
clearly $d_D\marker S_D\marker=v''_0\grammarstep[G']\cdots \grammarstep[G']
v''_n$ and $v''_n\in (T\cup \{S_\emptyset\})^*$.  Moreover, we have
$\Parikh{\pi_{C'\cup P'}(v''_n)}=d_D+\sum_{i=1}^n \xi_i$.

Let $g\colon X^*\to(T\cup \{S_\emptyset\})^*$ be the morphism with
$g(S)=S_\emptyset$ and that fixes the elements of $T$.  For a word $w\in
(N'\cup X)^*$ that contains $S_\emptyset$ and $1\le i\le n$, let $U_i(w)$ be
the word obtained from $w$ by replacing the first occurrence of $S_\emptyset$
by $g(u_i)$.  Then $w\grammarstep[G']U_i(w)$ and $\Parikh{\pi_{C'\cup
P'}(U_i(w))}=\Parikh{\pi_{C'\cup P'}(w)}+c_i$.  Thus, with
\[ u=U_n^{\nu(c_n)}\cdots U_1^{\nu(c_1)}(v''_n), \]
we have $v''_n\grammarsteps[G'] u\grammarsteps[G'] h(u)$ and hence $h(u)\in
\Lang{G'}$. By construction, $h(u)$ is in $M$ and thus $h(u)\in K'$. Moreover,
we have
\begin{align*}
\Parikh{\pi_{C'\cup P'}(h(u))}&=\Parikh{\pi_{C'\cup P'}(u)}=\Parikh{\pi_{C'\cup P'}(v''_n)}+\nu \\
&=d_D+\sum_{i=1}^n \xi_i+\nu=\mu.
\end{align*}
This proves $\bigcup_{c\in C'}c+P'^\oplus_c\subseteq \Parikh{\pi_{C'\cup P'}(K')}$.
\item \emph{Boundedness}. Let $w\in (C'\cup X\cup P'\cup \{\marker\})^*$ and $d_D\marker
S_D\marker\grammarsteps[G'] w$. By \cref{pa:one:boundedness} of our claim, we
have $|w|_{\marker}\le 2+|C|\cdot \ell$.
\item \emph{Insertion property}. Let $w\in (C'\cup X\cup P'\cup\{\marker\})^*$ and $d_D\marker
S_D\marker\grammarsteps[G'] w$. Then $\constantset{w}=\emptyset$ and
$h(\pi_{N'\cup X\cup\{\marker\}}(w))=\pi_{X\cup\{\marker\}}(w)$. Hence
\cref{pa:one:insertion} states that for each $\mu\in P'^\oplus_{d_D}$, there is
a $w'\in\SententialForms{G}$ with $\pi_{X\cup\{\marker\}}(w)\preceq_{\marker}
w'$ and $\Parikh{w'}=\Parikh{\pi_{X}(w)}+\varphi'(\mu)$.
\end{itemize}
\end{proof}

\begin{proof}[Proof of \cref{pa:onenonterminal}]
Let $G=(N,T,P,S)$. By \cref{pa:union}, we may assume that there is only one
production $S\to L$ in $P$.  By \cref{pa:checkif,pa:checkifnot}, one can
construct PAIM for $L_0=L\setminus (N\cup T)^*S(N\cup T)^*$ and for $L_1=L\cap
(N\cup T)^*S(N\cup T)^*$.

If $G'$ the grammar $G'=(N,T,P',S)$, where $P'=\{S\to L_1\}$ and $\sigma\colon
(N\cup T)^*\to\Powerset{(N\cup T)^*}$ is the substitution with $\sigma(S)=L_0$
and $\sigma(t)=\{t\}$ for $t\in T$, then
$\Lang{G}=\sigma(\SententialForms{G'})$.  Hence, one can construct a PAIM for
$\Lang{G}$ using \cref{pa:onenonterminal:sf,pa:substitution}.
\end{proof}

\section{Proof of \cref{pa:alg}}
\begin{proof}
Our algorithm works recursively with respect to the number of non-terminals.
In order to make the recursion work, we need the algorithm to work with right
hand sides in $\HG_i$. We show that, given $i\in\N$, an
$\HG_i$-grammar $G$, along with a PAIM in $\HG_i$ for each right
hand side in $G$, we can construct a PAIM for $L(G)$ in $\HG_i$.  A PAIM
for a language $L$ in $\HF_i$ can easily be turned into a PAIM for $L$ in
$\HG_i$.  Therefore, this statement implies the \lcnamecref{pa:alg}.

Let $G=(N,T,P,S)$ be an $\HG_i$-grammar and $n=|N|$.
For each $A\in N\setminus\{S\}$, let $G_A=(N\setminus\{S\},
T\cup\{S\}, P_A, A)$, where $P_A = \{ B\to L\in P \mid B\ne S\}$. Since $G_A$
has $n-1$ nonterminals, we can construct a PAIM for $\Lang{G_A}$ in
$\HG_i$ for each $A\in N\setminus\{S\}$.

Consider the substitution $\sigma\colon (N\cup T)^*\to\Powerset{(N\cup T)^*}$
with $\sigma(A)=\Lang{G_A}$ for $A\in N\setminus\{S\}$ and $\sigma(x)=\{x\}$
for $x\in T\cup\{S\}$.  Let $G'=(\{S\},T,P',S)$ be the $\HG_i$-grammar
with $P'=\{S\to\sigma(L) \mid S\to L\in P\}$.  By \cref{pa:substitution}, we
can construct a PAIM in $\HG_i$ for each right-hand-side of $G'$.
Therefore, \cref{pa:onenonterminal} provides a PAIM in $\HG_i$ for
$\Lang{G'}$. We claim that $\Lang{G'}=\Lang{G}$.

The inclusion $\Lang{G'}\subseteq\Lang{G}$ is easy to see: Each
$w\in\Lang{G_A}$ satisfies $A\grammarsteps[G] w$.  Hence, for $S\to L\in P$ and
$w\in\sigma(L)$, we have $S\grammarsteps[G] w$. This means
$\SententialForms{G'}\subseteq\SententialForms{G}$ and thus
$\Lang{G'}\subseteq\Lang{G}$.

Consider a derivation tree $t$ for $G$. We show by induction on the height of
$t$ that $\yield{t}\in\Lang{G'}$. We regard $t$ as a partial order.  A
\emph{cut} in $t$ is a maximal antichain.  We call a cut $C$ in $t$
\emph{special} if it does not contain the root, every node in $C$ has a label
in $T\cup \{S\}$, and if $x\in C$ and $y\le x$, then $y$ is the root or has a
label in $N\setminus\{S\}$.

There is a special cut in $t$: Start with the cut $C$ of all leaves. If there is a
node $x\in C$ and a non-root $y\le x$ with label $S$, then remove all nodes
$\ge y$ in $C$ and add $y$ instead. Repeat this process until it terminates.
Then $C$ is a special cut.

Let $u$ be the word spelled by the cut $C$. Since all non-root nodes $y<x$ for
some $x\in C$ have a label in $N\setminus\{S\}$, $u$ can be derived using a
production $S\to L$ once and then only productions $A\to M$ with $A\ne S$.
This means, however, that $u\in\sigma(L)$ and hence $S\grammarsteps[G'] u$.
The subtrees below the nodes in $C$ all have height strictly smaller than $t$.
Moreover, since all inner nodes in $C$ are labeled $S$, these subtrees are
derivation trees for $G$.  Therefore, by induction we have
$u\grammarsteps[G']\yield{t}$ and thus $S\grammarsteps[G']\yield{t}$.
\end{proof}

\section{Proof of \cref{pa:sli}}
\begin{proof}
According to \cref{pa:morphism}, it suffices to show that we can construct
a PAIM for $L\cap\ParikhInv{S}$.  Moreover, if $L=L_1\cup \cdots L_n$, then
\[L\cap\ParikhInv{S}=(L_1\cap\ParikhInv{S})\cup\cdots\cup (L_n\cap\ParikhInv{S}). \]
Thus, by \cref{pa:decomposelinear,pa:union}, we may assume that the PAIM for
$L$ is linear.  Let $(K,c,P,\varphi,\marker)$ be a linear PAIM for $L$ in
$\HG_i$.

The set $T = \{ \mu\in P^\oplus \mid \varphi(c+\mu) \in S\}$ is semilinear as
well, hence $T=\bigcup_{i=1}^n T_i$ for linear $T_i\subseteq P^\oplus$.
Write $T_i=\mu_i+F_i^\oplus$ with $\mu_i\in P^\oplus$, and $F_i\subseteq P^\oplus$ being
a finite set.  Let $P'_i$ be an alphabet with new symbols in bijection with the
set $F_i$ and let $\psi_i\colon P'^\oplus_i\to P^\oplus$ be the morphism
extending this bijection.  Moreover, let $U_i$ be the linear set
\[ U_i=\mu_i+\{p+\psi_i(p)\mid p\in P'_i\}^\oplus+(X\cup \{\marker\})^\oplus \]
and let $R_i=p_1^*\cdots p_m^*$, where $P'_i=\{p_1,\ldots,p_m\}$.  We claim
that with new symbols $c'_i$ for $1\le i\le n$, $C'=\{c'_i\mid 1\le i\le
n\}$, $P'=\bigcup_{i=1}^n P'_i$ and
\begin{align*}
\varphi'(c'_i)&=\varphi(c)+\varphi(\mu_i),      && \\
\varphi'(p)&=\varphi(\psi_i(p))                   &   & \text{for $p\in P'_i$}, \\
K'&=\bigcup_{i=1}^n c'_i\pi_{C'\cup X\cup P'\cup\{\marker\}}\left(c^{-1}KR_i\cap \ParikhInv{U_i}\right), 
\end{align*}
the tuple $(K',C',P',(P'_i)_{c'_i\in C'},\varphi',\marker)$ is a PAIM for
$L\cap\ParikhInv{S}$.

\begin{itemize}
\item\emph{Projection property} For $w\in L\cap\ParikhInv{S}$, we find a $cv\in
K$ with $\pi_X(cv)=w$.  Then $\varphi(\pi_{\{c\}\cup P}(cv))=\Parikh{w}\in S$
and hence $\Parikh{\pi_{P}(v)}\in T$. Let $\Parikh{\pi_P(v)}=\mu_i+\nu$ with
$\nu\in F_i^\oplus$, $P'_i=\{p_1,\ldots,p_m\}$, and $\psi_i(\kappa)=\nu$.
Then the word
\[ v'=vp_1^{\kappa(p_1)}\cdots p_m^{\kappa(p_m)} \]
is in $c^{-1}KR_i\cap\ParikhInv{U_i}$ and satisfies $\pi_X(v')=\pi_X(v)=w$.
Moreover, $v''=c'_i\pi_{C'\cup X\cup P'\cup\{\marker\}}(v')\in K'$ and hence $w=\pi_X(v'')\in
\pi_X(K')$. This proves $L\cap\ParikhInv{S}\subseteq\pi_X(K')$.

We clearly have $\pi_X(K')\subseteq\pi_X(K)=L$. Thus, it suffices to show
$\Parikh{\pi_X(K')}\subseteq S$.  Let $w=c'_iv\in K'$. Then $v=\pi_{C'\cup
X\cup P'\cup \{\marker\}}(v')$ for some $v'\in c^{-1}KR_i\cap\ParikhInv{U_i}$.
Let $P'_i=\{p_1,\ldots,p_m\}$ and write $v'=v''p_1^{\kappa(p_1)}\cdots
p_m^{\kappa(p_m)}$ for $\kappa\in P'^\oplus_i$.  This means $cv''\in K$ and
thus $\Parikh{\pi_X(cv'')}=\varphi(\pi_{\{c\}\cup P}(cv''))$ by the counting
property of $K$.  Since $v'\in\ParikhInv{U_i}$, we have
$\Parikh{\pi_P(cv'')}=\Parikh{\pi_{P}(v')}=\mu_i+\psi_i(\kappa)\in T_i$. Thus
\begin{align*}
\Parikh{\pi_X(w)}&=\Parikh{\pi_X(c'_iv)}=\Parikh{\pi_X(v')}=\Parikh{\pi_X(cv'')} \\
&=\varphi(\pi_{\{c\}\cup P}(cv''))\in \varphi(c+T_i)\in S.
\end{align*}
\item\emph{Counting property}
Let $w=c'_iv\in K'$ with $v=\pi_{C'\cup X\cup P'\cup \{\marker\}}(v')$ for some
$v'\in c^{-1}KR_i\cap\ParikhInv{U_i}$.  By definition of $U_i$, this implies
\[ \pi_{P}(v')=\mu_i + \psi_i(\pi_{P'}(v')) \]
and hence
\[ \varphi(\pi_P(v'))=\varphi(\mu_i)+\varphi(\psi_i(\pi_{P'}(v')))=\varphi(\mu_i)+\varphi'(\pi_{P'}(v')). \]
Moreover, if we write $v'=v''r$ with $cv''\in K$ and $r\in R_i$, then
\begin{align*}
\varphi'(\pi_{C'\cup P'}(w))&=\varphi'(c'_i)+\varphi'(\pi_{P'}(v')) \\
&=\varphi(c)+\varphi(\mu_i)+\varphi'(\pi_{P'}(v')) \\
&=\varphi(c)+\varphi(\pi_P(v'))=\varphi(\pi_{C\cup P}(cv'')) \\
&=\Parikh{\pi_X(cv'')}=\Parikh{\pi_X(w)}.
\end{align*}
This proves the counting property.
\item\emph{Commutative projection property}.
Let $\mu\in c'_i+P'^\oplus_i$, $\mu=c'_i+\kappa$ with $\kappa\in P'^\oplus_i$.
Let $P'_i=\{p_1,\ldots,p_m\}$. Then $\nu=\psi_i(\kappa)\in P^\oplus$ and the
commutative projection property of $K$ yields a $cv\in K$ with $\Parikh{\pi_{C\cup
P}(cv)}=c+\mu_i+\nu$. This means that the word
\[ v'=vp_1^{\kappa(p_1)}\cdots p_m^{\kappa(p_m)} \]
is in $c^{-1}KR_i\cap\Parikh{U_i}$. Furthermore, $\Parikh{\pi_{P'}(v')}=\kappa$ and hence
\begin{align*}
\Parikh{\pi_{C'\cup P'}(c'_i\pi_{C'\cup X\cup P'\cup\{\marker\}}(v'))}=c'_i+\kappa=\mu.
\end{align*}
This proves $\bigcup_{i=1}^n c'_i+P'^\oplus_i\subseteq\Parikh{\pi_{C'\cup P'}(K')}$.
The other inclusion follows directly from the definition of $K'$.
\item\emph{Boundedness} Since
$\pi_{\{\marker\}}(K')\subseteq\pi_{\{\marker\}}(K)$, $K'$ inherits boundedness
from $K$.
\item\emph{Insertion property} Let $c'_iw\in K'$ and $\mu\in P'^\oplus_i$. Write $w=\pi_{C'\cup X\cup P'\cup \{\marker\}}(v)$
for some $v\in c^{-1}KR_i\cap\ParikhInv{U_i}$, and $v=v'r$ for some $r\in R_i$. Then $cv'\in K$ and applying the insertion property of $K$ to
$cv'$ and $\psi_i(\mu)\in P^\oplus$ yields a $v''\in L$ with
$\pi_{X\cup\{\marker\}}(cv')\preceq_{\marker} v''$ and
$\Parikh{v''}=\Parikh{\pi_X(cv')}+\varphi(\psi_i(\mu))$.
This word satisfies
\begin{align*}
\pi_{X\cup\{\marker\}}(c'_iw)&=\pi_{X\cup\{\marker\}}(v)=\pi_{X\cup\{\marker\}}(cv')\preceq_{\marker} v'', \\
\Parikh{\pi_X(v'')}&=\Parikh{\pi_X(cv')}+\varphi(\psi_i(\mu)) \\
&=\Parikh{\pi_X(c'_iw)}+\varphi(\psi_i(\mu))=\Parikh{\pi_X(c'_iw)}+\varphi'(\mu).
\end{align*}
and it remains to be shown that $v''\in L\cap\ParikhInv{S}$. Since $v''\in L$,
this amounts to showing $\Parikh{v''}\in S$.

Since $\Parikh{v'}\in U_i$, we have $\Parikh{\pi_P(v')}\in \mu_i+F_i^\oplus$
and $\psi_i(\mu)\in F_i^\oplus$ and hence also
$\Parikh{\pi_P(v')}+\psi_i(\mu)\in \mu_i+F_i^\oplus=T_i$. Therefore,
\begin{align*}
\Parikh{v''}&=\Parikh{\pi_X(cv')}+\varphi(\psi_i(\mu)) \\
&=\varphi(\pi_{C\cup P}(cv'))+\varphi(\psi_i(\mu)) \\
&=\varphi(\pi_{C\cup P}(cv')+\psi_i(\mu))\in\varphi(c+T_i)\subseteq S.
\end{align*}
\end{itemize}
\end{proof}

\section{Proof of \cref{dc:overapprox}}

First, we need a simple auxiliary lemma.  For $\alpha,\beta\in X^\oplus$, we
write $\alpha\le \beta$ if $\alpha(x)\le\beta(x)$ for all $x\in X$.  For a set
$S\subseteq X^\oplus$, we write $\Dclosure{S}=\{\mu\in X^\oplus \mid \exists
\nu\in S\colon \mu\le \nu\}$ and $\Uclosure{S}=\{\mu\in X^\oplus \mid \exists
\nu\in S\colon \nu\le\mu\}$. The set $S$ is called \emph{upward closed} if
$\Uclosure{S}=S$. 
\begin{clemma}\label{basics:dclosure:recognizable}
For a given semilinear set $S\subseteq X^\oplus$, the set $\ParikhInv{\Dclosure{S}}$
is an effectively computable regular language.
\end{clemma}
\begin{proof}
The set $S'=X^\oplus\setminus(\Dclosure{S})$ is Presburger-definable in terms
of $S$ and hence effectively semilinear.  Moreover, since $\le$ is a
well-quasi-ordering on $X^\oplus$, $S'$ has a finite set $F$ of minimal
elements. Again $F$ is Presburger-definable in terms of $S'$ and hence
computable. Since $S'$ is upward closed, we have $S'=\Uclosure{F}$.  Clearly,
given $\mu\in X^\oplus$, the language $R_\mu=\{w\in X^* \mid \mu\le\Parikh{w}
\}$ is an effectively computable regular language. Since
$w\in\ParikhInv{\Dclosure{S}}$ if and only if $w\notin
\ParikhInv{\Uclosure{F}}$, we have
$X^*\setminus\ParikhInv{\Dclosure{S}}=\bigcup_{\mu\in F}R_\mu$. Thus, we can
compute a finite automaton for the complement, $\ParikhInv{\Dclosure{S}}$.
\end{proof}

\begin{proof}[Proof of \cref{dc:overapprox}]
We use \cref{pa:parikhannotations} to construct a PAIM $(K,C,P,(P_c)_{c\in
C},\varphi,\marker)$ for $L$ in $\HG_i$.

For each $c\in C$, we construct the semilinear sets $S_c=\{\mu\in P_c^\oplus
\mid \varphi(c+\mu)\in S\}$. By \cref{basics:dclosure:recognizable}, we can
effectively construct a finite automaton for the language
\[R=\bigcup_{c\in C} c\left(\ParikhInv{\Dclosure{S_c}}\shuffle (X\cup \{\marker\})^*\right).\]
We claim that $L'=\pi_X\left(K\cap R\right)$ is in $\HG_i$ and satisfies
$L\cap\ParikhInv{S}\subseteq L'\subseteq \Dclosure{(L\cap\ParikhInv{S})}$.  The
latter clearly implies $\Dclosure{L'}=\Dclosure{(L\cap\ParikhInv{S})}$.  Since
$K\in\HG_i$ and $\HG_i$ is an effective full semi-AFL, we clearly have
$L'\in\HG_i$.

We begin with the inclusion $L\cap\ParikhInv{S}\subseteq L'$. Let $w\in
L\cap\ParikhInv{S}$. Then there is a word $cv\in K$, $c\in C$ with
$\pi_X(v)=w$. Since $\Parikh{w}\in S$, we have $\varphi(\Parikh{\pi_{C\cup
P}(cv)})=\Parikh{\pi_X(v)}=\Parikh{w}\in S$ and hence $\Parikh{\pi_{P}(v)}\in
S_c\subseteq \Dclosure{S_c}$. In particular, $cv\in R$ and thus $w=\pi_X(cv)\in
L'$. This proves $L\cap\ParikhInv{S}\subseteq L'$.

In order to show $L'\subseteq\Dclosure{(L\cap\ParikhInv{S})}$, suppose $w\in L'$.
Then there is a $cv\in K\cap R$ with $w=\pi_X(cv)$.  The fact that $cv\in R$ means that
$\Parikh{\pi_{P_c}(v)}\in \Dclosure{S_c}$ and hence there is a $\nu\in
P_c^\oplus$ with $\Parikh{\pi_{P_c}(v)}+\nu\in S_c$.  This means in particular
\begin{equation}\Parikh{\pi_X(cv)}+\varphi(\nu)=\varphi(\pi_{C\cup P}(cv))+\varphi(\nu)\in S.\label{dc:insertion:satisfies}\end{equation}
The insertion property of $(K,C,P,(P_c)_{c\in C},\varphi,\marker)$ allows us
to find a word $v'\in L$ such that
\begin{align}
& \Parikh{v'}=\Parikh{\pi_X(cv)}+\varphi(\nu), && \pi_{X\cup\{\marker\}}(cv)\preceq_{\marker} v'.\label{dc:insertionword}
\end{align}
Together with \cref{dc:insertion:satisfies}, the first part of \cref{dc:insertionword} implies that $\Parikh{v'}\in S$.
The second part of \cref{dc:insertionword} means in particular that
$w=\pi_X(cv)\preceq v'$. Thus, we have $w\preceq v'\in L\cap\ParikhInv{S}$ and
hence $w\in \Dclosure{(L\cap\ParikhInv{S})}$.
\end{proof}

\section{Proof of \cref{pa:parikhannotations}}
\begin{clemma}[Finite languages]\label{pa:finite}
Given $L$ in $\HF_0$, one can construct a PAIM for $L$ in $\HF_0$.
\end{clemma}
\begin{proof}
Let $L=\{w_1,\ldots,w_n\}\subseteq X^*$ and define $C=\{c_1,\ldots,c_n\}$ and
$P=P_c=\emptyset$, where the $c_i$ are new symbols.  Let $\varphi\colon (C\cup
P)^\oplus\to X^\oplus$ be the morphism with $\varphi(c_i)=\Parikh{w_i}$. It
is easily verified that with $K=\{c_1w_1,\ldots,c_nw_n\}$, the tuple
$(K,C,P,(P_c)_{c\in C},\varphi,\marker)$ is a PAIM for $L$ in $\HF_0$.
\end{proof}

\begin{proof}[Proof of \cref{pa:parikhannotations}]
We compute the PAIM for $L$ recursively:
\begin{itemize}
\item If $L\in\HF_0$, we can construct a PAIM for $L$ in $\HF_0$ using \cref{pa:finite}.
\item If $L\in\HF_i$ and $i\ge 1$, then $L=h(L'\cap\ParikhInv{S})$ for some
$L'\subseteq X^*$ in $\HG_{i-1}$, a semilinear $S\subseteq X^\oplus$, and
a morphism $h\colon X^*\to Y^*$. We compute a PAIM for
$L'$ in $\HG_{i-1}$ and then use \cref{pa:sli} to construct a PAIM for $L$.
\item If $L\in\HG_i$, then $L=\Lang{G}$ for an $\HF_{i}$-grammar $G$. We
construct PAIM for the right-hand-sides of $G$ and then using \cref{pa:alg}, we
construct a PAIM for $L$ in $\HG_i$.
\end{itemize}
\end{proof}

\section{Proof of \cref{strictness:sli}}
\begin{proof}[Proof of \cref{strictness:sli}]
We write $Y=X\cup \{\#\}$. Suppose $(L\#)^*\in\HomSLI{\C}$. Then
$(L\#)^*=h(L'\cap\ParikhInv{S})$ for some $L'\subseteq Z^*$, a semilinear
$S\subseteq Z^\oplus$, and a morphism $h\colon Z^*\to Y^*$. Since $\C$ has
PAIMs, there is a PAIM $(K,C,P,(P_c)_{c\in C},\varphi,\marker)$ for $L'$ in
$\C$. Let $S_c=\{\mu\in P_c^\oplus \mid \varphi(c+\mu)\in S\}$.  Moreover, let
$g$ be the morphism with
\begin{align*}
g\colon (C\cup Z\cup P\cup \{\marker\})^*&\longrightarrow (Y\cup\{\marker\})^* && \\
z&\longmapsto h(z) & &\text{for $z\in Z$}, \\
x&\longmapsto \emptyWord    & &\text{for $x\in C\cup P$}, \\
\marker&\longmapsto \marker. & &
\end{align*}
Finally, we need the rational transduction $T\subseteq X^*\times (Y\cup
\{\marker\})^*$ with
\[ T(M)=\{s\in X^* \mid r\#s\#t\in M~\text{for some $r,t\in (Y\cup\{\marker\})^*$} \}. \]
We claim that
\begin{align*}
L=T(\hat{L}), && \text{where} && \hat{L}=\{ g(cw)\mid c\in C,~cw\in K,~\pi_P(w)\in \ParikhInv{\Dclosure{S_c}}\}.
\end{align*}
According to \cref{basics:dclosure:recognizable}, the language
$\ParikhInv{\Dclosure{S_c}}$ is regular, meaning $\hat{L}\in\C$ and hence
$T(\hat{L})\in\C$. Thus, proving $L=T(\hat{L})$ establishes the
\lcnamecref{strictness:sli}.

We begin with the inclusion $T(\hat{L})\subseteq L$. Let $s\in T(\hat{L})$ and
hence $r\#s\#t=g(cw)$ for $r,t\in (Y\cup\{\marker\})^*$, $c\in C$, $cw\in K$ and
$\pi_P(w)\in\ParikhInv{\Dclosure{S_c}}$. The latter means there is a $\mu\in
P_c^\oplus$ such that $\Parikh{\pi_P(w)}+\mu\in S_c$ and hence
\[\Parikh{\pi_{Z}(cw)}+\varphi(\mu)=\varphi(c+\Parikh{\pi_P(w)}+\mu)\in S.\]
By the insertion property of $K$, there is a $v\in L'$ with
$\pi_{Z\cup\{\marker\}}(cw)\preceq_{\marker} v$ and
$\Parikh{v}=\Parikh{\pi_Z(cw)}+\varphi(\mu)$. This means $\Parikh{v}\in S$ and
thus $v\in L'\cap\ParikhInv{S}$ and hence $g(v)=h(v)\in (L\#)^*$. Since
$g(\marker)=\marker$, the relation $\pi_{Z\cup\{\marker\}}(cw)\preceq_{\marker}
v$ implies
\[ r\#s\#t=g(cw)=g(\pi_{Z\cup\{\marker\}}(cw))\preceq_{\marker} g(v)\in (L\#)^*. \]
However, $\marker$ does not occur in $s$, meaning $\#s\#\in \#X^*\#$ is a factor of
$g(v)\in (L\#)^*$ and hence $s\in L$. This proves $T(\hat{L})\subseteq L$.

In order to show $L\subseteq T(\hat{L})$, suppose $s\in L$. The boundedness
property of $K$ means there is a bound $k\in\N$ with $|w|_{\marker}\le k$ for
every $w\in K$.  Consider the word $v=(s\#)^{k+2}$.  Since $v\in (L\#)^*$, we
find a $v'\in L'\cap\ParikhInv{S}$ with $v=h(v')$.  This, in turn, means there
is a $cw\in K$ with $c\in C$ and $\pi_Z(cw)=v'$. Then
\[ \varphi(c+\Parikh{\pi_P(w)})=\varphi(\pi_{C\cup P}(cw))=\Parikh{\pi_Z(cw)}=\Parikh{v'}\in S \]
and hence $\Parikh{\pi_P(w)}\in S_c\subseteq\Dclosure{S_c}$. Therefore,
$g(cw)\in\hat{L}\subseteq (Y\cup\{\marker\})^*$. Note that $g$ agrees with
$h(\pi_Z(\cdot))$ on all symbols but $\marker$, which is fixed by the former
and erased by the latter.  Since $h(\pi_Z(cw))=h(v')=v=(s\#)^{k+2}$, the word
$g(cw)$ is obtained from $(s\#)^{k+1}$ by inserting occurrences of $\marker$.
In fact, it is obtained by inserting at most $k$ of them since
$|g(cw)|_{\marker}=|cw|_{\marker}\le k$.  This means $g(cw)$ has at least one
factor $\#s\#\in \#X^*\#$ and hence $s\in T(g(cw))\subseteq T(\hat{L})$.  This
completes the proof of $L=T(\hat{L})$ and thus of the
\lcnamecref{strictness:sli}.
\end{proof}

\section{Proof of \cref{strictness:bursting}}
\begin{proof}
Suppose $G=(N,T,P,S)$ is $k$-bursting.  Let $\sigma\colon (N\cup T)^*\to
\Powerset{T^*}$ be the substitution with  $\sigma(x)=\{w\in T^{\le k} \mid
x\grammarsteps[G] w\}$ for $x\in N\cup T$. Since $\sigma(x)$ is finite for each
$x\in N\cup T$, there is clearly a locally finite rational transduction $T$
with $T(M)=\sigma(M)$ for every language $M\subseteq (N\cup T)^*$. In
particular, $\sigma(M)\in\C$ whenever $M\in\C$. Let $R\subseteq N$ be the set
of reachable nonterminals. We claim that
\begin{align} \Lang{G}\cap T^{>k}=\bigcup_{A\in R}\bigcup_{A\to L\in P} \sigma(L)\cap T^{>k}.\label{strictness:bursting:eq}\end{align}
This clearly implies $\Lang{G}\cap T^{>k}\in\C$.  Furthermore, since $\C$ is a
union closed full semi-trio and thus closed under adding finite sets of words,
it even implies $\Lang{G}\in\C$ and hence the \lcnamecref{strictness:bursting}.

We start with the inclusion ``$\subseteq$''. Suppose $w\in \Lang{G}\cap T^{>k}$
and let $t$ be a derivation tree for $G$ with $\yield{t}=w$. Since $|w|>k$, $t$
clearly has at least one node $x$ with $|\yield{x}|>k$. Let $y$ be maximal
among these nodes (i.e. such that no descendent of $y$ has a yield of length
$>k$).  Since $G$ is $k$-bursting, this means $\yield{y}=w$.  Furthermore, each
child $c$ of $y$ has $|\yield{c}|\le k$. Thus, if $A$ is the label of $y$, then
$A$ is reachable and there is a production $A\to L$ with $w\in \sigma(L)$.
Hence, $w$ is contained in the right-hand side of
\labelcref{strictness:bursting:eq}.

In order to show ``$\supseteq$'' of \labelcref{strictness:bursting:eq}, suppose
$w\in\sigma(L)\cap T^{>k}$ for some $A\to L\in P$ and a reachable $A\in N$. By
the definition of $\sigma$, we have $A\grammarsteps[G] w$.  Since $A$ is
reachable, there is a derivation tree $t$ for $G$ with an $A$-labeled node $x$
such that $\yield{x}=w$.  Since $G$ is $k$-bursting and $|w|>k$, this implies
$w=\yield{x}=\yield{t}\in \Lang{G}$ and thus $w\in\Lang{G}\cap T^{>k}$.
\end{proof}

\section{Proof of \cref{strictness:shuffle}}
\begin{proof}[Proof of \cref{strictness:shuffle}]
Let $K=L\shuffle \{a^nb^nc^n\mid n\ge 0\}$. If $K\in\Alg{\C}$, then also $M=K\cap
a^*(bX)^*c^*\in \Alg{\C}$. Hence, let $M=\Lang{G}$ for a reduced
$\C$-grammar $G=(N,T,P,S)$.  This means $T=X\cup \{a,b,c\}$. Let
$\alpha,\beta\colon T^*\to\Z$ be the morphisms with
\begin{align*} \alpha(w)=|w|_a-|w|_b, && \beta(w)=|w|_b-|w|_c.\end{align*}
Then $\alpha(w)=\beta(w)=0$ for each $w\in M\subseteq K$. Thus,
\cref{nonterminal:extension} provides $G$-compatible extensions
$\hat{\alpha},\hat{\beta}\colon (N\cup T)^*\to\Z$ of $\alpha$ and $\beta$,
respectively.

Let $k=\max\{|\hat{\alpha}(A)|, |\hat{\beta}(A)| \mid A\in N\}+1$ and consider the
$\C$-grammar $G'=(N,X,P',S)$, where $P'=\{ A\to\pi_{N\cup X}(L) \mid A\to L\in
P\}$. Then clearly $\Lang{G'}=\pi_X(M)=L$. We claim that $G'$ is $k$-bursting.
By \cref{strictness:bursting}, this implies $L=\Lang{G'}\in\C$ and hence the
\lcnamecref{strictness:shuffle}.

Let $t$ be a derivation tree for $G'$ and $x$ a node in $t$ with
$|\yield{x}|>k$.  Then by definition of $G'$, then there is a derivation tree
$\bar{t}$ for $G$ such that $t$ is obtained from $\bar{t}$ by deleting or
replacing by an $\emptyWord$-leaf each $\{a,b,c\}$-labeled leaf.
Since $x$ has to be an inner node, it has a corresponding node $\bar{x}$ in
$\bar{t}$. Since $G$ generates $M$, we have
\[ \yield{\bar{t}}=a^nbx_1bx_2\cdots bx_nc^n \]
for some $n\ge 0$ and $x_1,\ldots,x_n\in X$, $x_1\cdots x_n\in L$.  Moreover,
$\yield{\bar{x}}$ is a factor of $\yield{\bar{t}}$ and
$\pi_X(\yield{\bar{x}})=\yield{x}$. This means $|\pi_X(\yield{\bar{x}})|>k$ and
since in $\yield{\bar{t}}$, between any two consecutive $X$-symbols, there is a
$b$, this implies $|\yield{\bar{x}}|_{b} > k-1$. Let $A$ be the label of $x$
and $\bar{x}$.  By the choice of $k$, we have
$|\hat{\alpha}(\yield{\bar{x}})|=|\hat{\alpha}(A)|\le k-1$ and
$|\hat{\beta}(\yield{\bar{x}})|=|\hat{\beta}(A)|\le k-1$. Hence,
$|\yield{\bar{x}}|_b>k-1$ implies $|\yield{\bar{x}}|_a \ge 1$ and
$|\yield{\bar{x}}|_c\ge 1$.  However, a factor of $\yield{\bar{t}}$ that
contains an $a$ and a $c$ has to comprise all of $bx_1\cdots bx_n$.  Hence
\[ \yield{x}=\pi_X(\yield{\bar{x}})=x_1\cdots x_n=\pi_X(\yield{\bar{t}})=\yield{t}. \]
This proves that $G'$ is $k$-bursting.
\end{proof}

\section{Proof of \cref{strictness:hierarchy}}
\begin{proof}[Proof of \cref{strictness:hierarchy}]
First, note that if $V_i\in\HG_i\setminus\HF_i$, then
$U_{i+1}\in\HF_{i+1}\setminus\HG_i$: By construction of $U_{i+1}$, the fact
that $V_i\in\HG_i$ implies $U_{i+1}\in\HomSLI{\HG_i}=\HF_{i+1}$. By
\cref{hierarchy:closure}, $\HF_i$ is a union closed full semi-trio. Thus, if we
had $U_{i+1}\in \HG_i=\Alg{\HF_i}$, then \cref{strictness:shuffle} would imply
$V_i\in\HF_i$, which is not the case.

Second, observe that $U_{i+1}\in\HF_{i+1}\setminus\HG_i$ implies
$V_{i+1}\in\HG_{i+1}\setminus\HF_{i+1}$: By construction of $V_{i+1}$, the fact
that $U_{i+1}\in\HF_{i+1}$ implies $V_{i+1}\in\Alg{\HF_{i+1}}=\HG_{i+1}$.  By
\cref{hierarchy:closure}, $\HG_i$ is a full semi-AFL and by
\cref{pa:parikhannotations}, every language in $\HG_i$ has a PAIM in $\HG_i$.
Hence, if we had $V_{i+1}\in\HF_{i+1}=\HomSLI{\HG_i}$, then
\cref{strictness:sli} would imply $U_{i+1}\in\HG_i$, which is not the case.

Hence, it remains to be shown that $V_0\in\HG_0\setminus\HF_0$.  That, however,
is clear because $V_0=\#_0^*$, which is context-free and infinite.
\end{proof}

\end{document}